\documentclass[prb,aps,onecolumn,amsfonts,showpacs,superscriptaddress]{revtex4}

\usepackage{amsmath}
\usepackage{amsfonts}
\usepackage{amssymb}
\usepackage{graphicx}
\usepackage{epstopdf}
\usepackage{epsfig}
\usepackage{psfrag}
\usepackage{color}
\usepackage{subfigure}

\begin{document}

\title{Magnetothermoelectric Response near Quantum Critical Points} 

\author{M. J. Bhaseen} \affiliation{Theory of Condensed Matter Group,
Cavendish Laboratory, Cambridge, CB3 OHE, UK.}  \author{A. G. Green}
\affiliation{School of Physics and Astronomy, University of St
Andrews, North Haugh, St Andrews, Fife, KY16 9XP, UK.}
\author{S. L. Sondhi} \affiliation{Department of Physics, Princeton
University, Princeton, NJ 08544, USA.}  \date{\today}
 
\begin{abstract}
Following on from our previous work [Phys. Rev. Lett. {\bf 98}, 166801
(2007)] we examine the finite temperature magnetothermoelectric
response in the vicinity of a quantum critical point (QCP).  We begin
with general scaling considerations relevant to an arbitrary QCP,
either with or without Lorentz invariance, and in arbitrary
dimension. In view of the broad connections to high temperature
superconductivity, and cold atomic gases, we focus on the quantum
critical fluctuations of the relativistic Landau--Ginzburg
theory. This paradigmatic model arises in many contexts, and describes
the (particle-hole symmetric) superfluid--Mott insulator quantum phase
transition in the Bose--Hubbard model. The application of a magnetic
field opens up a wide range of physical observables, and we present a
detailed overview of the charge and thermal transport and
thermodynamic response. We combine several different approaches
including the epsilon expansion and associated Quantum Boltzmann
Equation (QBE), entropy drift, and arguments based on Lorentz
invariance. The results differ markedly from the zero field case, and
we include an extended discussion of the finite thermal conductivity
which emerges in the presence of a magnetic field.  We derive an
integral equation that governs its response and explore the crossover
upon changing the magnetic field. This equation may be interpreted as
a projection equation in the low field limit, and clearly highlights
the important r\^ole of collision invariants (or zero modes) in the
hydrodynamic regime.  Using an epsilon expansion around
three-dimensions, our analytic and numerical results interpolate
between our previously published value and the exact limit of
two-dimensional relativistic magnetohydrodynamics.

\end{abstract}

\pacs{73.43.Nq, 72.20.Pa, 74.25.Fy}

\maketitle 

\section{Introduction}
Quantum phase transitions (QPTs) in strongly correlated systems play
an important r\^ole in modern condensed matter physics. In problems
ranging from high temperature superconductivity to cold atomic gases,
dramatic changes in the ground state and physical response may often
accompany relatively small changes in the doping, the interaction
strength, or other system
parameters.\cite{Sachdev:QPT,Sondhi:Continuous} With the discovery of
high temperature superconductivity in the
cuprates,\cite{Bednorz:Possible} quantum phase transitions between
Mott-insulators (MIs) and superfluids (SFs) or superconductors, have
been at the forefront of this intense scrutiny.  More recently,
remarkable advances in cold atomic gases, have allowed the observation
of such transitions in systems of bosonic atoms.\cite{Greiner:SI} An
important stimulus for these studies, is that the SF-MI transition
separates two of the most fascinating phases of highly correlated
matter. The superfluid reveals the importance of phase coherence on
the macroscopic scale, and the Mott-insulator the importance of strong
interactions. The transition between the two clearly involves an
interplay between strong interactions and strong quantum
fluctuations. Such strongly correlated regimes are notoriously
difficult to analyze theoretically, and shedding light on this
enigmatic transition remains challenging.

\begin{figure}
\psfrag{KT}{2D KT}
\psfrag{INS}{Mott Insulator}
\psfrag{SF}{Superfluid}
\psfrag{N}{Normal}
\psfrag{T}{T}
\psfrag{g}{$g$}
\psfrag{c}{$g_c$}
\psfrag{XY}{$2+1$ XY QCP}
\includegraphics[width=6cm]{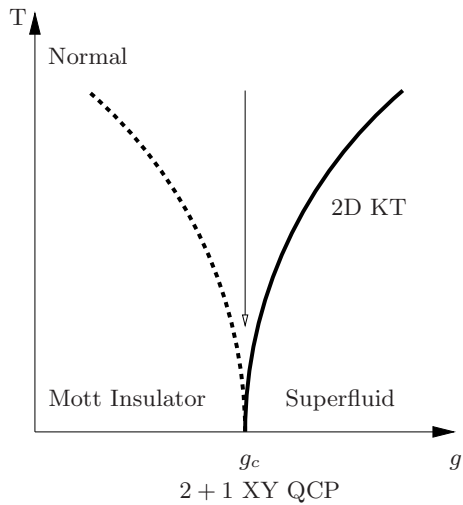}
\caption{Superfluid--Mott insulator quantum phase transition in $2+1$
dimensions, as tuned by a control parameter $g$, such as film
thickness or doping. The diagram shows the finite temperature 2D
Kosterlitz--Thouless transition, and the crossover between the Mott
insulator and the normal state. We examine the universal finite
temperature magnetothermoelectric response in the vicinity of the
(particle--hole symmetric) $2+1$ XY QPT, as shown by the vertical
arrow.}
\label{Fig:sfmikt}
\end{figure}

In this work, we focus on the magnetothermoelectric response in the
vicinity of such quantum critical points (QCPs). Our motivation for
this, and our previous work,\cite{Bhaseen:magneto} arose in connection
with high temperature superconductivity, where experiments indicate
strong superconducting fluctuations in a broad range of temperatures
above ${\rm T}_{\rm c}$.\cite{Xu:Vortex} These signatures appear in
both thermodynamics and transport measurements performed in magnetic
fields, and include enhanced diamagnetism,\cite{Wang:Dia} and a strong
Nernst signal.\cite{Wang:Nernst} The Nernst response is the transverse
electric field induced by a thermal gradient in a magnetic field, and
therefore hinges on the interplay of several different probes.
Although a tremendous amount of theoretical progress has been made in
various regions of the phase
diagram,\cite{Ussishkin:Gaussian,Ussishkin:Diag,Oganesyan:Nernst,
Uss:Interp,Mukerjee:Nernst,Oganesyan:Dia,Podolsky:Nernst}
much less was known about the complete magnetothermoelectric response
in the vicinity of such QCPs.\cite{Uss:Critical} In view of the
enhanced fluctuations, and the prospect of universal results, we
advocated examining this problem at a simple, but rather generic SF-MI
transition in the XY universality class\cite{Bhaseen:magneto} --- see
Fig~\ref{Fig:sfmikt}. Somewhat more ambitiously, we set out with a
view to describe the full complement of magnetothermoelectric response
coefficients.

Our strategy\cite{Bhaseen:magneto} is to focus on the particle--hole
symmetric transitions in the ubiquitous Bose--Hubbard model. This
model has a distinguished history, and provides a paradigmatic example
of a SF-MI quantum phase
transition.\cite{Fisher:Bosonloc,Fisher:Presence,Cha:Universal,
Damle:Nonzero,Damle:Thesis,Green:Nonlinear,Green:current}
While it arose primarily in connection with bosonic models of strongly
correlated {\em electron} systems, where the bosons are to be thought
of as Cooper pairs, it has since been cleanly realized in ground
breaking experiments in cold atomic Bose gases.\cite{Greiner:SI}
Although not directly linked to a fermionic high temperature
superconductor (with a d-wave order parameter and nodal
quasiparticles) the simplicity of the Bose--Hubbard model is
appealing. Amongst its many virtues, it admits a description as a
quantum Landau--Ginzburg theory (or Abelian Higgs model) and so it
naturally embraces ${\rm U}(1)$ phase fluctuations. Such fluctuations
have long been argued to play an important r\^ole in high temperature
superconductors.\cite{Fisher:Thermal,Emery:Imp} More recently, the
finite temperature {\em classical fluctuations} of the
Kosterlitz--Thouless\cite{KT:KT} transition have been argued to
influence their diamagnetic
response.\cite{Oganesyan:Dia,Podolsky:Nernst} In this work we focus on
the vicinity of the QCP, and study the impact of {\em quantum critical
fluctuations} on the fundamental transport coefficients, and other
physical response functions --- see Fig.~\ref{Fig:sfmikt}. In contrast
to other approaches, which tackle related problems from the superfluid
side, and thus in terms of vortices, it is quite natural to examine
the critical region in terms of particle--hole excitations of the Mott
insulator. Although dual vortex formulations are possible, we do not
pursue this complementary approach here.

From a condensed matter perspective, our primary interests are in the
charge and thermal transport, and the thermodynamic response in the
vicinity of the QCP. This critical point is of course well studied and
there is a vast literature on its zero field properties which we do
not attempt to review.  In order to place our work in
context,\cite{Bhaseen:magneto} we simply recollect the most recent
precursors pertaining to single field {\em transport} measurements in
the absence of a magnetic field. It was recognized in early works that
the electrical conductivity is a finite universal multiple of
$e^2/h$.\cite{Fisher:Presence,Fisher:Bosonloc,
Cha:Universal,Damle:Nonzero,Damle:Thesis,Sorensen:Universal,Green:Nonlinear}
Interactions at the fixed point are essential in order to render this
finite, as opposed to a more conventional, non-interacting single
carrier Drude peak. More specifically, this may be traced to
collisions between counter propagating particles and holes, and in
general requires a finite frequency, and finite temperature,
hydrodynamic treatment of the
problem.\cite{Damle:Nonzero,Damle:Thesis} In contrast, it was well
understood that the thermal conductivity {\em diverges} at the clean
interacting fixed point; \cite{Senthil:Un,Vojta:RG,Green:Nonlinear} in
the presence of a thermal gradient, particle and hole excitations move
in the {\em same} direction, and the collision mechanism is unable to
render this quantity finite. In the absence of a magnetic field, a
finite thermal response therefore requires the introduction of
impurities, or other forms of energy relaxation.\cite{Green:Nonlinear}

As we noted in our previous work,\cite{Bhaseen:magneto} this
circumstance is changed markedly in the presence of a magnetic
field. Although the DC Hall conductivity vanishes on symmetry grounds
at the particle-hole symmetric point, the application of a magnetic
field opens up the possibility of non-vanishing thermoelectric
response coefficients, {\em even in the absence of impurities, or
other forms of scattering}. For this reason, we confined ourselves to
the clean case, in order to better expose the main universal
results. This is not a serious limitation, since results in the
presence of impurities may be obtained from the finite frequency
behavior of the clean system, provided it's not driven to a new
non-trivial fixed point. The general problem one is interested in, is
thus to apply various combinations of $E$, $B$, $\nabla T$, possibly
at finite frequency, and to measure the associated response
coefficients. In view of the conceptual importance of the
Bose--Hubbard model and the XY universality class, we present a
variety of approaches to the general magnetothermoelectric response.

The layout of this paper is as follows: In section \ref{scaling} we
begin with a general discussion of scaling close to a QCP. This
overview is relevant to both bosonic and fermionic systems in
arbitrary dimension, and thus helps to correlate the more detailed,
model specific results we shall present. In section \ref{FT} we recall
the field theory approach to the superfluid--insulator transition in
the Bose--Hubbard model, and the simplifications at the particle-hole
symmetric point. In section \ref{BE} we describe the Boltzmann
approach to quantum critical
transport.\cite{Damle:Nonzero,Damle:Thesis} In sections \ref{EFR},
\ref{TGR} and \ref{MFR}, we present a brief overview of the properties
in electric fields, magnetic fields, and temperature gradients taken
{\em separately}. In section \ref{sect:crossed} we examine the
behavior in combined crossed electric and magnetic fields, and discuss
two distinct regimes of behavior. We approach this problem in several
different ways, including entropy transport, Lorentz invariance, and a
linear response analysis of the QBE. In section \ref{Sect:TGM} we
examine the behavior in the presence of a temperature gradient and a
magnetic field, and once again discuss two regimes. We verify that the
Onsager relations are satisfied and obtain a non-vanishing thermal
conductivity. We conclude in section \ref{sect:conc} and provide
several technical appendices.

Whilst this longer manuscript was under construction, elegant
extensions of this work appeared which also include the effects of
impurities and particle-hole symmetry breaking at relativistic
QCPs.\cite{Hartnoll:Nernst} These reveal important links between the
transport coefficients, and develop connections to the high energy
community.\cite{Herzog:Quantum} Applications to other gapless systems
such as graphene have also been
investigated.\cite{Fritz:Clean,Muller:Collective,Muller:Graphene}

\section{Scaling Forms}
\label{scaling}
As usual, the approach to a continuous phase transition is accompanied
by a divergent correlation length, $\xi$, and a divergent correlation
time, $\xi_\tau\sim \xi^{z}$, where $z$ is the so-called dynamical
exponent. Close to the transition, dependence on the microscopic
details drops out, and non-trivial scaling relations between physical
observables and the system parameters may be obtained. In the case of
a {\em quantum} phase transition, where the microscopic energy scale
is tuned to zero, the only relevant energy scale in the problem is the
temperature, $T$. In this case, the divergent correlation time is
given by
\begin{equation}
\xi_\tau\sim 1/T,
\label{corrtime}
\end{equation}
where we set, $\hbar=k_B=1$, for simplicity; see for example the
review \cite{Sondhi:Continuous} for more details.  This diverging time
scale is accompanied by a divergent correlation length
\begin{equation}
\xi\sim (\xi_t)^{1/z}\sim T^{-1/z},
\label{corrlength}
\end{equation}
where a non-trivial dynamical exponent, $z$, reflects the potential
for disparity in the temporal and static correlations.  Throughout
this work will be interested in the magnetic, electrical and thermal
response in the vicinity of a QCP. Here, we examine the general
consequences which follow from simple, yet powerful, scaling ideas.
As exemplified by measurements on quantum Hall systems, such
considerations are able to correlate a wide variety of different
physical probes, yet are crucially independent of the microscopic
approach employed.\cite{Sondhi:Continuous}

The space-time dimensions of the electric and heat currents contribute
to the overall temperature dependence:
\begin{align}
[J_e] & \sim {\rm Time}^{-1}{\rm Length}^{-(d-1)} \sim T^{1+(d-1)/z},
\label{dimelcur}\\
[J_h] & \sim {\rm Time}^{-2}{\rm Length}^{-(d-1)}\sim T^{2+(d-1)/z}. 
\label{dimhcur}
\end{align}
In general, we are also interested in the dependence of these currents
on the measuring frequency and the external fields, and it is
temperature against which these are compared. Having pinned the
overall dimensions, these external fields will enter via dimensionless
ratios involving the temperature.\cite{Sondhi:Continuous} Using the
defining relations, ${\bf E}=\dot{\bf A}$, and 
${\bf B}=\nabla \times {\bf A}$,
\begin{align}
[{\bf E}] & \sim  {\rm Length}^{-1}{\rm Time}^{-1}\sim T^{1+1/z}, \\
[{\bf B}] & \sim {\rm Length}^{-2}\sim T^{2/z},
\end{align}
where $[{\bf A}]={\rm Length}^{-1}$. In this way we arrive at the
following scaling forms, valid in arbitrary dimension and for generic
dynamical exponent:
\begin{equation}
J_e(T,{\bf E},{\bf B},\nabla T,\omega) \sim
T^{1+(d-1)/z}\,F_e\left(\frac{|{\bf E}|}{T^{1+1/z}},\frac{|{\bf
    B}|}{T^{2/z}},\frac{|\nabla T|}{T^{1+1/z}},\frac{\omega}{T}\right),
\end{equation}
\begin{equation}
J_h(T,{\bf E},{\bf B},\nabla T,\omega) \sim
    T^{2+(d-1)/z}\,F_h\left(\frac{|{\bf E}|}{T^{1+1/z}},\frac{|{\bf
    B}|}{T^{2/z}},\frac{|\nabla
    T|}{T^{1+1/z}},\frac{\omega}{T}\right),
\label{heatscal}
\end{equation}
where, ${\rm F}_e$ and ${\rm F}_h$ are universality class dependent
scaling functions. One may readily incorporate additional
perturbations in a similar fashion.  As we discuss in section
\ref{MFR}, similar considerations also apply to thermodynamic
quantities obtained from the scaling form for the free energy.  We
emphasize that our only assumption in deriving these scaling forms is
proximity to a QCP. In particular, they are independent of the
statistics of the underlying carriers, and are equally valid for both
bosonic and fermionic systems. These expressions are invaluable as
they enable one to confirm, and sometimes infer, the field and
temperature dependence of the transport coefficients. Perhaps more
importantly, they also allow one to correlate a large number of
distinct scenarios and probes, without lengthy or sophisticated
computations.  For example, as we will discuss in section \ref{EFR},
in the absence of an applied magnetic field, or temperature gradient,
linear response in $E$ immediately yields
\begin{equation}
J_e(T,E,\omega) \sim T^{(d-2)/z}F(\omega/T)E\equiv \sigma(\omega,T)E.
\end{equation}
The general dependence of the AC conductivity on frequency and
temperature is therefore easily read
off.\cite{Damle:Nonzero,Damle:Thesis} Of course, in order to pin the
precise functional dependence on these variables, explicit
calculations of the scaling functions are necessary, and we turn our
attention to this problem below. Most crucially, at low frequencies,
$\omega\ll T$, collisions at the fixed point necessitate a
hydrodynamic, or quantum Boltzmann treatment of the critical
regime.\cite{Damle:Nonzero,Damle:Thesis} For simplicity, we consider
the, $z=1$, relativistic field theory approach to the (particle-hole
symmetric) SF-MI transition in the Bose--Hubbard model, but our
interests, and overall approach are clearly broader.

\section{Field Theory}
\label{FT}
\begin{figure}
\psfrag{MI}{Mott Insulator}
\psfrag{SF}{Superfluid}
\psfrag{m}{$\mu/U$}
\psfrag{t}{$t/U$}
\includegraphics[width=6cm]{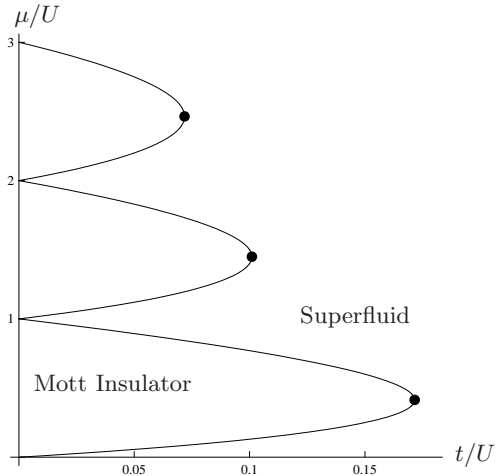}
\caption{Mean field phase diagram of the Bose--Hubbard model. At the
tips of the Mott insulating lobes (indicated by dots) the system is
particle--hole symmetric, and described by the relativistic quantum
Landau--Ginzburg theory.}
\label{Fig:bhlobes}
\end{figure}

The Bose--Hubbard model has received considerable attention in recent
years,\cite{Fisher:Bosonloc,Fisher:Presence,Cha:Universal,Damle:Nonzero,Damle:Thesis,Green:Nonlinear,Green:current}
and describes bosons hopping on a lattice with amplitude $t$, and
interacting via a short range repulsive interaction $U$:
\begin{equation}
H=-t\sum_{\langle ij\rangle}(b_i^\dagger b_j+b_j^\dagger
b_i)-\mu\sum_i n_i+\frac{U}{2}\sum_i n_i(n_i-1).
\label{bh}
\end{equation}
The Bose creation and annihilation operators satisfy the usual
commutation relations, $[b_i,b_j^\dagger]=\delta_{ij}$, where
$n_i=b_i^\dagger b_i$, is the number of bosons at site $i$, and $\mu$
is the chemical potential.  In the context of a Josephson array or
superconductor, the bosons represent Cooper pairs of charge $Q=2e$,
tunnelling between superconducting regions. In general, one may also
include the effects of disorder and long range interactions into such
a model, but here we shall concentrate on the simplest case
(\ref{bh}). The phase diagram of the Bose--Hubbard model is well
established, and exhibits both superfluid and Mott insulating
regions,\cite{Fisher:Bosonloc} the latter occurring for strong enough
repulsive interactions --- see Fig.\ref{Fig:bhlobes}. As a function of
the chemical potential, $\mu$, this model exhibits a series of Mott
insulating ``lobes'' where the density of bosons is pinned to
successive integers. At a given point within the Mott lobes, the
energy cost for producing particle (or hole) excitations is measured
by the vertical displacement to the upper (or lower) phase
boundary. At the tips of these lobes, the energy cost to producing
particle--hole excitations vanishes, and the model is particle--hole
symmetric. In addition, the density remains constant as one enters the
superfluid phase along a trajectory of constant chemical potential,
passing through this apex. In the vicinity of these points, the SF-MI
transition is described by the relativistic action of an interacting
complex scalar field $\Phi$, \cite{Fisher:Bosonloc}
\begin{equation}
S=\int d^Dx\,|\partial_\mu\Phi|^2-m^2|\Phi|^2-\frac{u_0}{3}|\Phi|^4,
\label{SLG}
\end{equation}
where $D=d+1$, and $d$ is the number of spatial dimensions of the
original Bose--Hubbard model (\ref{bh}), and the mass parameter, $m$,
is set by the temperature. This is nothing but a relativistic, quantum
Landau--Ginzburg action for the superconducting order parameter,
$\Phi$, and its associated fluctuations. It therefore represent a
useful starting point to unravel the more general problem of phase
fluctuations at SF-MI transitions. Away from these particle--hole
symmetric points, the density changes as one enters the superfluid
phase, and the action picks up an additional term, linear in the time
derivative. Correspondingly, the dynamical exponent changes from $z=1$
to $z=2$; see for example Ref~\onlinecite{Sachdev:QPT}. Here, we will
focus on the case with $z=1$, since it will allow us to employ the
powerful machinery of relativistic quantum field theory. More general
results, for arbitrary $z$, may be obtained by appealing to the
general scaling arguments outlined in section \ref{scaling}.

\section{Quantum Boltzmann Equation}
\label{BE}

A convenient and physically intuitive way to think about this problem,
at leading order in the epsilon expansion, is by means of the quantum
Boltzmann approach to quantum critical
transport.\cite{Damle:Nonzero,Damle:Thesis} In this framework we may
regard the model (\ref{SLG}) as a gas of particle--hole excitations of
the Mott insulator.  The quantum Boltzmann equation (QBE) emerges at
lowest order in the epsilon expansion and describes the impact of weak
scattering (as controlled by epsilon) on these quasiparticles at the
Gaussian fixed point. It takes the form of a nonlinear
integro-differential equation for the momentum space distribution
functions, $f_\pm({\bf k},t)$, of such particle and hole excitations
\begin{equation}
\frac{{\partial f}_{\pm}}{\partial t}\pm Q\left({{\bf E}+{\bf v}_{\bf
k}\times{\bf B}}\right).\frac{\partial f_\pm}{\partial{\bf k}}={\rm
I}_\pm[f_+,f_-],
\label{QBE}
\end{equation}
where ${\bf v}_{\bf k}\equiv \partial {\varepsilon}_{\bf k}/\partial
{\bf k}$ and $\varepsilon_{\bf k}=\sqrt{{\bf k}^2c^2+m^2c^4}$. For
simplicity we consider a spatially homogeneous system in uniform
external fields. The collision term represents scattering between
these excitations, and most crucially, incorporates the nonlinear
interaction of the Landau--Ginzburg field theory (\ref{SLG}) and the
associated critical fluctuations:\cite{Damle:Nonzero,Damle:Thesis}
\begin{eqnarray}
{\rm I}_\pm & = & -\frac{2u_0^2}{9}\int \prod_{i=1}^3 \frac{d^d{\bf
k}_i} {(2\pi)^d\,2\varepsilon_{k_i}}\left(\frac{{\mathcal F}_\pm^{\rm
out}-{\mathcal F}_\pm^{\rm in}}{2{\varepsilon}_k}\right)\times
\nonumber \\ & & \hspace{-1cm} (2\pi)^{d+1}\delta({\bf k}+{\bf
k}_1-{\bf k}_2-{\bf k}_3)\delta(\varepsilon+\varepsilon_1-
\varepsilon_2-\varepsilon_3),
\label{col}
\end{eqnarray}
where scattering out of state ${\bf k}$ is given by
\begin{eqnarray}
{\mathcal F}_\pm^{\rm out} & = & 2f_\pm({\bf k})f_\mp({\bf
k}_1)[1+{f}_\pm({\bf k}_2)][1+ {f}_\mp({\bf k}_3)] \nonumber \\ & & +
f_\pm({\bf k})f_\pm({\bf k}_1)[1+{f}_\pm({\bf k}_2)][1+{f}_\pm({\bf
k}_3)],\nonumber 
\end{eqnarray} 
and we have suppressed the explicit time dependence of the
distribution functions. Scattering in to state ${\bf k}$ follows by
interchanging $f_\pm$ and $1+f_\pm$. The structure of the collision
term (\ref{col}) is readily seen by using Fermi's Golden rule, and
represents the leading term in the $\epsilon$-expansion of the
associated Keldysh field theory.\cite{Damle:Thesis} The factors of
$1+f$ remind us that we are dealing with a system of bosons in this
example. As in standard treatments of $|\Phi|^4$ theory, it is
convenient to access the non-trivial fixed point by means of an
epsilon expansion around the upper critical
dimension.\cite{Fazio:Epsilon} At the Wilson--Fisher fixed point,
where the renormalized mass vanishes, the bare couplings must be tuned
to the values \cite{Damle:Nonzero}
\begin{equation}
m^2=\frac{4\pi^2T^2\epsilon}{15}, \quad u_0=\frac{24\pi^2\epsilon}{5},
\label{couplings}
\end{equation}
where our spatial dimensionality is given by $d=3-\epsilon$; since we
are primarily interested in two spatial dimensions, we shall set
$\epsilon=1$, at the end of any calculations. The collision term
(\ref{col}) is therefore proportional to $\epsilon^2$. As we shall
see, the leading order epsilon expansion is illuminating both from a
quantitative numerical perspective, and also in its ability to expose
the external field and temperature dependence of physical quantities.
This semiclassical approach is formally justified within the epsilon
expansion where, at the temperatures of interest, the mean free path is
much longer than the thermal de Broglie wavelength and the mean
interparticle spacing. Further information on the relativistic
QBE,\cite{Degroot:Relkin} and applications to the quark gluon plasma,
may also be found in the literature.\cite{Blaizot:Boltz}

In this Boltzmann representation, the ${\rm U}(1)$ electric current of
the field theory (\ref{SLG}) takes the form
\begin{eqnarray}
{\bf J}_e=Q \int \frac{d^d k}{(2\pi\hbar)^d}\,{\bf v}_{\bf
k}\,[f_+({\bf k},t)-f_-({\bf k},t)],
\label{bolel}
\end{eqnarray}
and the heat current is given (in linear response) by
\begin{equation}
{\bf J}_h=\int \frac{d^d k}{(2\pi\hbar)^d}\,{\bf v}_{\bf
k}\varepsilon_{\bf k}\,[f_+({\bf k},t)+f_-({\bf k},t)].
\label{bolheat}
\end{equation}
Here, we use ${\bf k}$ to label the momentum and not wavevector; the
dimensions of the currents are $[{\bf J}_e]=Cm^{-(d-1)}s^{-1}$ and
$[{\bf J}_h]=Jm^{-(d-1)}s^{-1}$. In this representation, the field
theory conservation laws are related by appropriate momentum space
integrals (or moments) of the QBE.

The strategy is thus the same as in any application of the Boltzmann
equation.  We should solve the appropriate QBE for the non-equilibrium
distribution function(s), and then compute the associated transport
currents.  We may then extract the fundamental transport coefficients
defined via
\begin{equation}
\begin{pmatrix} {\bf J}_e^{\rm tr} \\ {\bf J}_h^{\rm tr} \end{pmatrix} =
\begin{pmatrix}\sigma & \alpha \\ \tilde\alpha & \bar\kappa  
\end{pmatrix}\begin{pmatrix}{\bf E}\\ -\nabla T\end{pmatrix},
\label{tcoeffs}
\end{equation}
where $\tilde\alpha=T \alpha$ is required by the Onsager
relations,\cite{Abrikosov:Fund,Onsager:R1,Onsager:R2} and we denote
the response to a temperature gradient by an overbar\footnote{We are grateful to Subir Sachdev for suggesting this clearer notation.} in order to
distinguish it from the thermal conductivity defined in the absence of
a particle current.\footnote{In general it is
necessary to distinguish between the {\em microscopic} currents
(\ref{bolel}) and (\ref{bolheat}), and the {\em transport} currents
(\ref{tcoeffs}), due to the presence of magnetization
currents.\cite{Cooper:Thermo,Obraztsov:EMF} This issue is 
circumvented in conventional Boltzmann approaches to transport in 
magnetic fields which do not include magnetization corrections to 
the distribution function.} Knowledge of the transport coefficients is 
particularly useful since 
they allow one to correlate a wide variety of different measurements. 
For example, the Nernst coefficient, $\nu$, is the transverse electric field
generated under open circuit conditions by a temperature gradient and
a magnetic field. Imposing, ${\bf J}_e^{{\rm tr}}=0$, on equation
(\ref{tcoeffs}) one may relate this {\em open circuit} measurement to
the more primitive {\em transport} coefficients
by\cite{Ussishkin:Gaussian}
\begin{equation}
\nu\equiv \frac{1}{B}\frac{E_y}{(-\nabla T)_x}=
\frac{1}{B}\frac{\alpha_{xy}\sigma_{xx}-\alpha_{xx}\sigma_{xy}}
{\sigma_{xx}^2+\sigma_{xy}^2}.
\end{equation}
At the particle--hole symmetric point we have chosen to focus on, the
Hall coefficient, $\sigma_{xy}$, vanishes and this may be reduced to
\begin{equation}
\nu=\frac{1}{B}\frac{\alpha_{xy}}{\sigma_{xx}}.
\label{phnu}
\end{equation}
Such an approximate reduction may also occur in situations where
$\sigma_{xy}\approx 0$. We see that the transverse thermoelectric
response, $\alpha_{xy}$, is central to a discussion of the Nernst
response. We shall examine this quantity in detail from several
different perspectives, beginning in section \ref{sect:crossed}.  We
shall also discuss the behavior of the other transport coefficients,
including the thermal conductivity, $\bar\kappa_{xx}(B)$.

\section{Electric Field Response}
\label{EFR}
Before embarking on a detailed discussion of the general
magnetothermoelectric response, it is instructive to recall the main
known results pertaining to single field
measurements.\cite{Fisher:Bosonloc,Fisher:Presence,Damle:Nonzero} As
we discussed in section \ref{scaling}, in the absence of any applied
temperature gradients or magnetic fields, one expects the linear
response electric current to behave as
\begin{equation}
J_e(T,{\bf E},\omega)\sim T^{1+(d-1)/z}\frac{|{\bf E}|}{T^{1+1/z}}\,
\Sigma\left(\frac{\omega}{T}\right).
\end{equation}
The corresponding conductivity therefore reads
\begin{equation}
\sigma(T,\omega)\sim T^{(d-2)/z}\,\Sigma\left(\frac{\omega}{T}\right).
\label{sigscale}
\end{equation}
For $z=1$, this is borne out by the direct Boltzmann calculations of
Damle and Sachdev.\cite{Damle:Nonzero} Indeed, the nontrivial scaling
(\ref{sigscale}) as a function of $\omega/T$ was an important catalyst
for their Boltzmann treatment; in general the limits
$\omega\rightarrow 0$ and $T\rightarrow 0$ do not commute and
$\Sigma(0)\neq \Sigma(\infty)$. As such the DC response at finite
temperature differs from that obtained at zero temperature. The
universal scaling function, $\Sigma(\omega/T)$, may be computed
numerically in the low frequency hydrodynamic regime by means of the
QBE (\ref{QBE}), and at zero frequency one
obtains\cite{Damle:Nonzero,Damle:Thesis}
\begin{equation}
\Sigma(0)=\frac{0.1650}{\epsilon^2}.
\label{DSSig}
\end{equation}
In $d=2$, where $\epsilon=1$, the DC conductivity is a universal
multiple of $e^2/h$:
\begin{equation}
\sigma(0)=\Sigma(0)\frac{(2e)^2}{\hbar}\simeq 1.037 \frac{4e^2}{h}.
\end{equation}
This result is clearly appealing, and is remarkably close to both the
self-dual value,\cite{Fisher:Mag,Wen:Universal} $4e^2/h$, and to a
number of early experiments on thin films.\cite{Damle:Nonzero} More
generally, the order of limits is also relevant to the electric field
itself. A discussion of the non-commutativity of $E,\omega,T$, and the
associated non-linear response, may be found in
Refs.~\onlinecite{Green:Nonlinear,Green:current}.

It is worth noting that the result (\ref{DSSig}) is {\em singular} in
the small parameter $\epsilon$. This is a direct reflection of the
proximity to the Gaussian fixed point, and that collisions must
overcome the conventional Drude response for a single non-interacting
carrier. This is possible in this two carrier system because electric
current relaxation does not violate momentum
conservation.\cite{Damle:Nonzero} A steady state with a finite
conductivity is therefore established. As we shall see, the specific
leading order dependence on $\epsilon$ is intimately tied to the
physical quantity under consideration. For example, in section
\ref{sect:crossed} we shall demonstrate that the leading order
contribution to $\alpha_{xy}$ is {\em regular} in the epsilon
expansion. To leading order, one may therefore drop the ${\mathcal
O}(\epsilon^2)$ collision term in the QBE and consider bosons of mass
${\mathcal O}(\sqrt{\epsilon})$ moving under the action of the applied
fields. This is in stark contrast to the case considered by Damle and
Sachdev,\cite{Damle:Nonzero} where the collision term was crucial in
order to render the electrical conductivity finite and proportional to
$1/\epsilon^2$. Nonetheless, as we shall discuss below, the collision
term is central to a better understanding of the thermal conductivity
in the drift regime. Here the leading order contribution begins at
${\mathcal O}(\epsilon^2)$ and is {\em inversely} related to the
electrical conductivity. For uniformity of presentation we shall
include the collision term throughout, and simplify when it is
appropriate.

\section{Temperature Gradient Response} 
\label{TGR}
In view of the finite DC electrical conductivity discussed above, it
is tempting to think that the system might also support a finite
thermal conductivity, in the spirit of a conventional Wiedemann--Franz
relation. A little reflection reveals that this is {\em not} the case
however, and that the thermal conductivity in fact diverges in the
clean system.\cite{Senthil:Un} A simple way to see this is to note
that under the action of the thermal gradient, particles and holes
move in the {\em same} direction. As such, relaxation of the
corresponding heat current requires {\em energy} relaxation, which is
not contained in either the original model (\ref{SLG}) or the
Boltzmann description (\ref{QBE}). (This is in contrast to the
electrical conductivity, where particles and holes move in {\em
opposite} directions, with no net momentum. Collisions are highly
effective in limiting the electrical current, without violating the
conservation laws.\cite{Damle:Nonzero}) It follows that the divergence
of the thermal conductivity is intimately tied to the conservation of
the energy momentum tensor,\cite{Senthil:Un} in much the same way as
happens in low-dimensional integrable systems.\cite{Shimshoni:Thermal}
As we shall discuss below, this conservation argument no longer
applies in the presence of a magnetic field.\cite{Bhaseen:magneto} By
minimally coupling the field theory (\ref{SLG}) to a magnetic field,
the conservation law is modified. As such, in the presence of a
magnetic field, a non-vanishing thermoelectric tensor and thermal
conductivity may be supported, {\em even in the clean homogeneous
system}.\cite{Bhaseen:magneto} Impurities and other scattering
mechanisms such as umklapp processes, may of course render these
quantities finite also, but in the first instance, it is clearly
essential to understand the universal results pertaining to the clean
homogeneous system. Before embarking on a general discussion of the
interplay between a magnetic field and other probes and response
functions, let us examine the response to a magnetic field alone.

\section{Magnetic Field Response}
\label{MFR}

In view of the interest in fluctuation
diamagnetism\cite{Oganesyan:Dia,Podolsky:Nernst} it is
instructive to investigate the response to a magnetic field
alone. Recalling our scaling arguments from section \ref{scaling}, in
the vicinity of a QCP, the only relevant energy scale against which to
compare the magnetic field is the temperature. In the absence of
electrical and thermal gradients, we therefore expect the free energy
density to scale as\cite{Bhaseen:magneto}
\begin{equation}
{\mathcal F}(T,B)\sim T^{1+d/z}\,f_1\left(\frac{B}{T^{2/z}}\right).
\label{freescal}
\end{equation}
Here we have used the fact that the correlation length, $\xi$,
diverges with the correlation time, $\xi_\tau\sim 1/T$, according to
$\xi\sim(\xi_\tau)^{1/z}$, and so the prefactor is an energy
density. In addition, we have used the fact that $[B]\sim 1/L^2\sim
T^{2/z}$, as follows from our discussion in section \ref{scaling}.  In
weak magnetic fields, $B\ll T^{2/z}$, we may expand this in powers of
$B$:
\begin{equation}
{\mathcal F}(T,B)\sim T^{1+(d-4)/z}B^2.
\end{equation}
Here we assume that symmetry under reversal of the magnetic field
ensures that only even powers of $B$ appear in the weak field
expansion of the scaling function $f_1$.  It follows that the linear
response magnetization scales as\cite{Bhaseen:magneto}
\begin{equation}
M=-\frac{\partial {\mathcal F}}{\partial B}\sim T^{1+(d-4)/z}B.
\end{equation}
This is consistent with a finite temperature, diagrammatic Kubo
calculation of the magnetic susceptibility of a charged scalar field,
with $m\sim T$ and $z=1$. On the other hand, in strong magnetic
fields, $B\gg T^{2/z}$, it is natural to expand this in a power series
in temperature, and recast the scaling relation (\ref{freescal}) in
the form
\begin{equation}
{\mathcal F}(T,B)\sim B^{(d+z)/2}{\tilde f}_1\left(\frac{T}{B^{z/2}}\right).
\end{equation}
In strong fields, or low temperatures, $B\gg T^{2/z}$, we thus expect
\begin{equation}
{\mathcal F}(T,B)\sim B^{(d+z)/2}.
\end{equation}
In the relativistic problem, this yields the strong field behavior,
${\mathcal F}\sim B^{3/2}$, in $d=2$, and ${\mathcal F}\sim B^2$, in
$d=3$. Broadly speaking, this non-trivial field dependence is a
reflection of the relativistic Landau level spectrum, where ${\mathcal
E}\sim \sqrt{B}$. This is suggested on dimensional grounds by equation
(\ref{freescal}) for $z=1$, and is borne out in a direct computation
of the partition function of a free massive relativistic charged
scalar field.\cite{Jana:Charged,Jana:Universal} These considerations
are also compatible with elegant and highly non-trivial results
obtained in the early days of quantum electrodynamics on vacuum
polarization and pair
production;\cite{Heis:Euler,Weisskopf:HE,Schwinger:Gauge} see for
example Ref. \onlinecite{Dunne:HE} for a recent review of
Heisenberg--Euler effective Lagrangians in both spinor and scalar
quantum electrodynamics.\footnote{Note that at {\em weak} fields, the
$B^2$ contributions to the free energy are eliminated in these works
by a renormalization of the electric charge associated with the
Maxwell term.}  These detailed studies require ultraviolet
regularization and yield an additional logarithm at strong magnetic
fields, so that ${\mathcal F}\sim B^2\ln(1/B)$ in $d=3$. Although this
is not captured by the simple scaling arguments, it is intimately
connected to the vacuum screening properties of the field theory via
the renormalization group beta function.

In this section we have discussed the distinct field regimes of the
static magnetic response. In the subsequent discussion, we shall see
how such regime divisions also emerge in the transport properties. In
many respects this is rather natural, since the dynamics and statics
are intimately related at QCPs.\cite{Sondhi:Continuous}

\section{Crossed Electric and Magnetic Fields}
\label{sect:crossed}
\begin{figure}
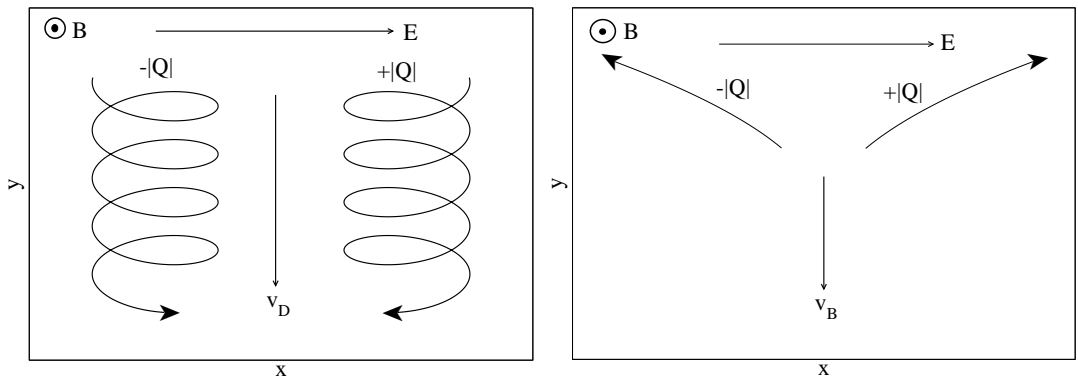

\subfigure{
\includegraphics[width=7cm,clip=true]{magnetolongfig3.eps}\label{fig:drift}}
\subfigure{
\includegraphics[width=7cm,clip=true]{magnetolongfig4.eps}\label{fig:egb}}
\caption{Motion of a single relativistic charged particle in crossed
electric and magnetic fields. \subref{fig:drift} In the regime $|{\bf
E}|<c|{\bf B}|$ the particle executes cyclotron orbits and has a well
defined transverse drift velocity $v_D$. \subref{fig:egb} In the
regime $|{\bf E}|>c|{\bf B}|$ the particle is continually accelerated
by the electric field.}
\label{Fig:Regimes}
\end{figure}

In order to understand the quantum critical transport in combined
${\bf E}$ and ${\bf B}$ fields, it is instructive to recall the motion
of a {\em single} relativistic particle in crossed electric and
magnetic fields.\cite{Jackson:Classical} This will provide significant
orientation for the more general {\em interacting} field
theory.\cite{Bhaseen:magneto} The most important feature is that the
motion of a single charged particle is qualitatively different
depending on whether, $E<c_0B$, or $E>c_0B$, where $c_0$ is the speed
of light. This is most easily understood from the vantage point of a
moving frame of reference.\cite{Jackson:Classical} In the former case,
there always exists a moving frame where the electric field vanishes,
and the particle experiences a pure magnetic field. Conversely, in the
latter case, there always exists a frame where the magnetic field
vanishes, and the particle experiences a pure electric field. Boosting
back to the lab frame, we mix in the complementary field component,
but the qualitative character of the motion is more ``electric field
like'' or ``magnetic field like'', as dictated by the inequalities ---
see Fig.~\ref{Fig:Regimes}. We shall discuss this in more detail
below, but before doing so, we note that our problem is a little more
subtle due to the appearance of an {\em effective speed of light},
$c$, in the effective field theory (\ref{SLG}), and the associated QBE
(\ref{QBE}). What matters for our purposes, at least in the first
instance, is the solutions to the QBE (\ref{QBE}) for a given fixed
ratio of $E$ and $B$. It may be seen that these solutions are
kinematically distinct for, $E<cB$, and $E>cB$, as follows directly
from the left hand side of the differential equation, without recourse
to Lorentz invariance arguments. As, such it the {\em effective}
Lorentz structure which plays a key r\^ole in determining the
solutions of the relativistic QBE (\ref{QBE}), and we shall henceforth
use this effective speed of light in our subsequent discussions. It is
interesting to note that such Lorentz transformation arguments also
find applications in other effective relativistic systems, as was
recently discussed by Lukose {\em et al} in the context of
graphene.\cite{Lukose:Graphene}

Under a Lorentz transformation with a velocity ${\bf v}$, the electric
and magnetic fields transform according to \cite{Jackson:Classical}
\begin{eqnarray}
{\bf E}^\prime & = & \gamma ({\bf E}+{\bf v}\times{\bf
B})-\frac{\gamma^2}{\gamma+1}\frac{{\bf v}(\bf{v}.{\bf E})}{c^2}, \\
{\bf B}^\prime & = & \gamma\left({\bf B}-\frac{{\bf v}\times{\bf
E}}{c^2}\right)-\frac{\gamma^2}{\gamma+1}\frac{{\bf v}({\bf v}.{\bf
B})}{c^2},
\end{eqnarray}
where, $\gamma=1/\sqrt{1-v^2/c^2}$, and we work in SI units. It is
readily seen that for crossed ${\bf E}$ and ${\bf B}$ fields, with
$|{\bf E}|<c|{\bf B}|$, there exists a frame moving at the drift
velocity
\begin{equation}
{\bf v}_{\rm D}\equiv \frac{{\bf E}\times{\bf B}}{|{\bf B}|^2},
\label{vdrift}
\end{equation}
where the electric field ${\bf E}^\prime$ vanishes. In this moving
frame, the particle is subject to a pure magnetic field of reduced
strength, ${\bf B}^\prime={\bf B}/\gamma_{\rm D}$. Taking our magnetic
field to point along the z-axis, the particle executes cyclotron
orbits in the moving frame with, $x^\prime(t^\prime)=r\cos
(\omega^\prime t^\prime)$ and $y^\prime(t^\prime)=\mp
r\sin(\omega^\prime t^\prime)$; here $r$ is the radius of the orbit,
$\omega^\prime$ is the cyclotron frequency, and the signs indicate the
sense of rotation for positive and negative charges respectively. With
the electric field along the $x$-axis, the drift velocity points along
the negative $y$-axis. Boosting back to the lab frame using the
inverse Lorentz transformations one finds the parametric equations of
motion:
\begin{equation}
x(t)=r\cos(\omega^\prime t^\prime), \quad 
y(t)=\gamma(\mp r\sin(\omega^\prime t^\prime)-v_Dt^\prime),
\quad t=\gamma(t^\prime \pm v_{\rm D}r\sin(\omega^\prime t^\prime)/c^2).
\end{equation}
In the nonrelativistic limit $c\rightarrow\infty$, these reduce to the
parametric equations of a trochoid: $x^2(t)+(y(t)+v_{\rm D}t)^2=r^2$
--- see Fig~\ref{Fig:Regimes}. In view of the cyclotron motion of
equal numbers of particles and holes, it follows that the DC
conductivity, $\sigma_{xx}(B)=0$, in this regime, at least at the
single particle level. On the other hand we see that the
thermoelectric tensor, $\alpha_{xy}$, may be finite due to the
finiteness of the drift velocity. We shall see that these expectations
are borne out, even in the presence of interactions at the fixed
point, and that this single particle description captures the relevant
physics.\cite{Bhaseen:magneto} It also underpins the divergence of the
Nernst coefficient (\ref{phnu}) in the clean, particle-hole symmetric
case.\cite{Bhaseen:magneto}

Returning to our single particle problem, for crossed ${\bf E}$ and
${\bf B}$ fields with $|{\bf E}|>c|{\bf B}|$, there exists a frame
moving at velocity
\begin{equation}
{\bf v}_{\rm B}\equiv c^2\left(\frac{{\bf E}\times{\bf B}}{|{\bf E}|^2}\right),
\end{equation}
where the magnetic field ${\bf B}^\prime$ vanishes. In this moving
frame, the charged particle is subject to a pure electric field of
reduced strength, ${\bf E}^\prime={\bf E}/\gamma_{\rm B}$. In the
absence of scattering, the energy $\varepsilon^\prime$, and the
components of momentum ${\bf p}^\prime$ parallel to ${\bf E}$,
continue to increase indefinitely. In the lab frame
\begin{eqnarray}
\varepsilon & = &\gamma_{\rm B}(\varepsilon^\prime+{\bf v}_{\rm B}.{\bf p}^\prime), 
\label{elab} \\
{\bf p}_{\parallel} & = & \gamma_{\rm B}({\bf p}_\parallel^\prime +
{\bf v}_{\rm B}\varepsilon^\prime/c^2),\label{ppar}\\
{\bf p}_\perp & = & {\bf p}_\perp^\prime,
\label{pperp}
\end{eqnarray}
where the labels parallel and perpendicular are with respect to the
{\em boost velocity}, ${\bf v}_{\rm B}$. In these notations ${\bf
p}_\parallel^\prime$ is constant (since it is transverse to the
electric field) whilst ${\bf p}_\perp^\prime$ and
$\varepsilon^\prime\equiv\sqrt{{c^2{{\bf p}^\prime}^2+m^2c^4}}$
increase with time.  It follows from equations (\ref{elab}),
(\ref{ppar}) and (\ref{pperp}) that the energy, and {\em both}
components of the momentum increase indefinitely in the lab
frame. That is to say, if we apply an electric field $|{\bf E}|>c|{\bf
B}|$, both particles and holes acquire an identical and ever
increasing component of the momentum at right angles to the electric
field.\footnote{We are extremely grateful for illuminating
conversations with David Huse on this point.} This {\em cannot} be
relaxed by the collision term. Since, ${\varepsilon}_k {\bf
v}_k=c^2{\bf k}$, we expect that $\alpha_{xy}$ is infinite in this
regime. On the other hand, compatibility with the results of Damle and
Sachdev, for ${\bf B}=0$, suggests the possibility of a finite value
of $\sigma_{xx}$. Once again, the underlying divergence of the Nernst
coefficient (\ref{phnu}) is apparent.

From the preceeding discussion, we see how the ratio, $E/B$, may
influence transport measurements. As we shall discuss in section
\ref{Sect:TGM}, there are analogous regimes in a thermal gradient, for
$\nabla T\lesssim B$, and $\nabla T\gtrsim B$, at least within the
framework of the collisionless Boltzmann equation with a linearized
driving term. Although we no longer have Lorentz invariance arguments,
the distinction once again shows up in the single particle kinematics
of the associated QBE, since the thermal gradient acts like a
momentum-dependent electric field. In the remainder of this section,
we shall take the electric field regimes in turn, and examine the
magnetothermoelectric transport coefficients from a variety of
different perspectives. We will focus primarily on the drift regime,
$|{\bf E}|<c|{\bf B}|,$ since it is both the simplest to analyze, and
also pertains to conventional linear response measurements at fixed
magnetic field.

\subsection{Drift Regime: $|{\bf E}|<c|{\bf B}|$}
In this section we are interested in the thermoelectric response in
the drift regime. It turns out that there are several complementary
ways to address this problem, and we examine these below. Our strategy
is to first develop an understanding of the QBE (\ref{QBE}) as
written, and to defer discussion of the epsilon expansion itself until
later.  This combination of different perspectives is particularly
useful in establishing the Onsager relations between the transport
coefficients.\cite{Abrikosov:Fund,Onsager:R1,Onsager:R2} It also helps
demonstrate the equivalence between the field theory and quantum
Boltzmann approaches.  We begin in section \ref{Sect:entropy} with a
computation of the thermoelectric tensor, $\alpha_{xy}$, based on the
entropy drift of a charged scalar field. In section \ref{Sect:BE} we
turn our attention to the more general problem of the Lorentz
invariant solutions of the QBE. In section \ref{Sect:THC} we use this
explicit distribution to determine the heat current response to an
electric field, and verify the answer obtained by entropy drift. In
section \ref{Sect:LR} we demonstrate how this result also follows from
a more familiar linearization of the QBE. Such a linearization
approach will be particularly useful in the presence of thermal
gradients, where Lorentz field transformation arguments are not
available. In section \ref{Sect:Num} we finally turn to the epsilon
expansion itself, and evaluate our general expression for
$\alpha_{xy}$ numerically. We compare the results to our scaling
analysis of section \ref{scaling}. We provide a very brief discussion
of the non-drift regime in section \ref{Sect:CDR}.

\subsubsection{Entropy Flow}
\label{Sect:entropy}
As follows from the defining relations (\ref{tcoeffs}), the
thermoelectric tensor, $\alpha$, quantifies the electric current which
flows in response to a temperature gradient. Provided the Onsager
relations are satisfied, this may also be obtained (up to a factor of
temperature) from the heat current which flows in response to a
temperature gradient. The latter route is easier to begin with, and we
consider the complementary approach in section \ref{Sect:TGM}.  In
view of the well defined drift velocity it is natural to compute the
transverse thermoelectric response, $\alpha_{xy}$, as a transport of
entropy with the characteristic flow\begin{equation} {\bf v}_{\rm
D}=\frac{{\bf E}\times{\bf B}}{|{\bf B}|^2}.
\end{equation}
To lowest order in the epsilon expansion, it is sufficient to compute
the entropy density of a free massive charged scalar field, where the
mass parameter is given by (\ref{couplings}). That is to say, the mass
incorporates the leading Hartree contribution of the
self-interactions.\cite{Damle:Nonzero} A free boson is essentially a
harmonic oscillator, and the partition function of the latter is
readily seen to be
 \begin{equation}
Z=\sum_{n=0}^\infty e^{-\beta(n+1/2)\hbar\omega}=
\frac{1}{2\sinh\left(\frac{\beta\hbar \omega}{2}\right)}.
\end{equation}
Taking the logarithm and integrating over all momenta, the analogous
expression for a massive {\em neutral} scalar field follows
immediately:\cite{Kapusta:Finite}
\begin{equation}
\ln Z = - V \int \frac{d^dk}{(2\pi\hbar)^d}\left(\frac{\beta\hbar\omega_k}{2}
+\ln(1-e^{-\beta\hbar\omega_k})\right),
\end{equation}
where $\hbar\omega_k\equiv
\varepsilon_k=\sqrt{k^2c^2+m^2c^4}$. Dropping the first (divergent
zero point) contribution, the free energy density reads
\begin{equation}
{\mathcal F}=k_BT\int \frac{d^dk}{(2\pi\hbar)^d}\,\ln\left(1-e^{-\beta\varepsilon_k}
\right).
\end{equation}
The corresponding entropy density is given by
\begin{equation}
{\mathcal S}=-\frac{\partial{\mathcal F}}{\partial T}=
-k_B\int \frac{d^dk}{(2\pi\hbar)^d}\,\left[\ln\left(1-e^{-\beta\varepsilon_k}\right)
-\frac{\beta\varepsilon_k}{e^{\beta\varepsilon_k}-1}\right].
\end{equation}
Performing an integration by parts on the logarithmic term one obtains
\begin{equation}
{\mathcal S} = \frac{1}{dT}\int \frac{d^dk}{(2\pi\hbar)^d}\, 
f_0(\varepsilon_k)\,\nabla_{\bf k}.(\varepsilon_k{\bf k}),
\end{equation}
where $f_0=(e^{\beta\varepsilon_k}-1)^{-1}$ is the Bose distribution
function, and we have used the identity $\nabla_{\bf
k}.(\varepsilon_k{\bf k})={\bf v}_{\bf k}.{\bf k}+d\varepsilon_k$.
Since particles and holes acquire the {\em same} drift velocity, the
corresponding heat current density for our charged scalar field theory
is given by ${\bf J}_h=2T{\mathcal S}{\bf v}_{\rm D}$. In particular,
if we apply an electric field, $E_x$, in the positive $x$-direction,
and a magnetic field, $B_z\equiv B$, in the positive $z$-direction,
the drift velocity points in the {\em negative} $y$-direction:
\begin{equation}
J_h^y=-2T{\mathcal S}\frac{E_x}{B_z}.
\end{equation}
Assuming the validity of the Onsager relations, which we demonstrate
are satisfied in section \ref{Sect:TGM}, we thus obtain
\begin{equation}
\alpha_{yx}=-\frac{2{\mathcal S}}{B}.
\end{equation}
Equivalently, using the Onsager symmetry relation\cite{Abrikosov:Fund}
$\alpha_{xy}({\bf B})=\alpha_{yx}(-{\bf B})$
\begin{equation}
\alpha_{xy}=\frac{2{\mathcal S}}{B}=\frac{2}{dBT}\int 
\frac{d^dk}{(2\pi\hbar)^d}\,f_0(\varepsilon_k)\nabla_{\bf k}.(\varepsilon_k{\bf k}).
\end{equation}
A useful alternative form of this result, which better exposes the
relation to heat currents, is obtained by performing another
integration by parts:
\begin{equation}
\alpha_{xy}=\frac{2}{dBT}\int 
\frac{d^dk}{(2\pi\hbar)^d}\,\varepsilon_k{\bf k}.\left(-\nabla_{{\bf k}}f_0\right).
\end{equation}
Using the identity
\begin{equation}
\nabla_{\bf k}f_0=\frac{c^2{\bf k}}{\varepsilon_k}
\left(\frac{\partial f_0}{\partial\varepsilon_k}\right),
\end{equation}
we may also write this in scalar form as
\begin{equation}
\alpha_{xy}=\frac{2{\mathcal S}}{B}=\frac{2c^2}{dBT}\int \frac{d^dk}{(2\pi\hbar)^d}\, 
k^2\left(-\frac{\partial f_0}{\partial \varepsilon_k}\right).
\label{enalpha}
\end{equation}
Note that the result for $\alpha_{xy}$ is positive in sign; an
electric field in the $y$-direction and a magnetic field in
$z$-direction yields ${\bf E}\times{\bf B}$ drift, and thus a
transport of heat, in the positive $x$-direction.  In the next
sections we shall see how this result (\ref{enalpha}) emerges from the
QBE (\ref{QBE}), {\em even in the presence of
interactions}. Developing the Boltzmann approach is particularly
useful since it will allow an intuitive and systematic calculation of
all the transport coefficients in this regime. Before closing the
section let us note that the entropy density per species of carrier
has a particularly simple form in the {\em massless} limit. With $m=0$
one obtains
\begin{equation}
{\mathcal S}={\mathcal C}_d \left(k_B\lambda_T^{-d}\right),
\end{equation}
where $\lambda_T\equiv \hbar c/k_B T$ is a thermal wavelength for
massless particles, and
\begin{equation}
{\mathcal C}_d\equiv \frac{1}{d}\int \frac{d^d{\bar k}}{(2\pi)^d}\frac{\bar k^2 e^{\bar k}}{(e^{\bar k}-1)^2},
\end{equation}
where $\bar k\equiv c k/k_B T$ are dimensionless variables. In
particular, ${\mathcal C}_3=2\pi^2/45\approx 0.439$, and ${\mathcal
C}_2=3\zeta(3)/2\pi\approx 0.574$. We shall employ the first of these
results within the epsilon expansion in section \ref{Sect:Num}.

\subsubsection{Lorentz Invariance and the QBE}
\label{Sect:BE}

The first step in developing the Boltzmann approach is to solve the
QBE for the non-equilibrium distribution functions, $f_\pm({\bf
k},t)$.  In the absence of any electric and magnetic fields, it is
readily verified that the QBE (\ref{QBE}), is satisfied by the static
Bose distribution function
\begin{equation}
f_\pm({\bf k},t)=f_0(\varepsilon_{\bf k})\equiv\frac{1}{e^{\beta\varepsilon_{\bf k}}-1},
\label{bdist}
\end{equation}
{\em even in the presence of the collision term}. This is easily seen
by noting that $1+f_0(\varepsilon_{\bf k})=e^{\beta\varepsilon_{\bf
k}}f_0(\varepsilon_{\bf k})$. It follows that ${\mathcal F}_\pm^{\rm
out}$ equals ${\mathcal F}_\pm^{\rm in}$ provided $\varepsilon({\bf
k})+\varepsilon({\bf k}_1)=\varepsilon({\bf k}_2)+\varepsilon({\bf
k}_3)$. The energy conserving delta function present in equation
(\ref{col}) is therefore enough to ensure that the collision term
vanishes and that the full QBE is satisfied.

Let us now examine the non-equilibrium situation in the presence of
crossed electric and magnetic fields. As discussed in section
\ref{sect:crossed}, in the regime $|{\bf E}|<c|{\bf B}|$, we may move
to a frame with velocity
\begin{equation}
{\bf v}_{\rm D}=\frac{{\bf E}\times {\bf B}}{|{\bf B}|^2},
\label{vd}
\end{equation}
where the electric field vanishes. Since a pure magnetic field does
not affect the energy of a charged particle, it follows than an
equilibrium distribution $f_0(\varepsilon^\prime_{\bf k})$ must
satisfy the Boltzmann equation in this boosted frame. (Indeed, it is
readily verified by direct substitution that $f_\pm({\bf
k})=f_0(\varepsilon_{\bf k})$ is a solution of the original QBE
(\ref{QBE}) when ${\bf E}=0$ and ${\bf B}\neq 0$, since $\partial
f_0/\partial{\bf k}={\bf v}_k\partial f_0/\partial \varepsilon_{\bf
k}$.) Since
\begin{equation}
{\varepsilon}_{\bf k}^\prime=\gamma_{\rm D}(\varepsilon_{\bf k}-{\bf
v}_{\rm D}.{\bf k}),
\end{equation}
where, $\gamma_{\rm D}=(1-{\bf v}_{\rm D}^2/c^2)^{-1/2}$, we conclude that 
\begin{equation}
f_\pm({\bf k})=f_0(\varepsilon^\prime_{\bf
k})=f_0\left(\frac{{\varepsilon}_{\bf k}-{\bf v}_{\rm D}.{\bf
k}}{\sqrt{1-{\bf v}_{\rm D}^2/c^2}}\right),
\label{boostdist}
\end{equation}
is a solution of the full QBE (\ref{QBE}). Again, it may be verified by
direct substitution that (\ref{boostdist}) is in fact a solution of
(\ref{QBE}) in the presence of our crossed ${\bf E}$ and ${\bf B}$
fields. Explicitly, the left hand side of the Boltzmann equation is
readily seen to give zero using the fact that ${\bf B}\times{\bf
v}_{\rm D}={\bf E}$:
\begin{equation}
({\bf E}+{\bf v}_{\bf k}\times{\bf B}).\frac{\partial
f_{\pm}}{\partial {\bf k}}=({\bf E}+{\bf v}_{\bf k}\times{\bf
B}).({\bf v}_{\bf k}-{\bf v}_{\rm D})\,\gamma_{\rm D}\frac{\partial
f_0(\varepsilon_{\bf k}^\prime)}{\partial \varepsilon_{\bf
k}^\prime}\propto{\bf v}_{\bf k}.({\bf E}-{\bf B}\times{\bf v}_{\rm
D})=0.
\end{equation}
Likewise, in the collision term we see that ${\mathcal F}^{\rm
out}_\pm$ equals ${\mathcal F}^{\rm in}_\pm$ provided that
$\varepsilon^\prime({\bf k})+\varepsilon^\prime({\bf
k}_1)=\varepsilon^\prime({\bf k}_2)+\varepsilon^\prime({\bf
k}_3)$. Since $\varepsilon^\prime({\bf k})=\gamma_{\rm
D}(\varepsilon_{\bf k}-{\bf v}_{\rm D}.{\bf k})$, this is ensured by
the combined action of the energy and momentum conserving delta
functions appearing in (\ref{col}). In this way we have established
the non equilibrium solution (\ref{boostdist}) of the QBE valid in the
drift regime $|{\bf E}|<c|{\bf B}|$. This is analogous to Kohn's
theorem\cite{Kohn:Cyclo} for non-relativistic electron
systems.\cite{Schofield:PC} In the next section we shall use this
distribution to compute the transport coefficients.

\subsubsection{Transverse heat current in response to an electric field}
\label{Sect:THC}
The equality of the particle and hole distribution functions
(\ref{boostdist}) reflects the fact that drift velocity (\ref{vd}) is
independent of the charge of the carrier. It follows immediately from
the definition (\ref{bolel}), that the both the longitudinal and the
transverse components of the electrical conductivity vanish in this
limit. Note that there is no conflict with the existing results of
Damle and Sachdev,\cite{Damle:Nonzero,Damle:Thesis} since our present
results are derived in the drift regime $|{\bf E}|<c|{\bf B}|$. As
such we cannot simply set ${\bf B}=0$ and recover the results obtained
in the absence of ${\bf B}$. Moreover, the vanishing of the
conductivities are consistent with the single particle picture
presented in Fig.~\ref{Fig:Regimes}. In contrast, it is clear from the
definition (\ref{bolheat}), that a non-vanishing heat current may be
supported in crossed ${\bf E}$ and ${\bf B}$ fields. Substituting the
drift solution (\ref{boostdist}) into (\ref{bolheat}) yields
\begin{equation}
{\bf J}_h=2\int \frac{d^dk}{(2\pi\hbar)^d}\,c^2{\bf k}
f_0\left(\frac{{\varepsilon}_{\bf k}-{\bf v}_{\rm D}.{\bf
k}}{\sqrt{1-{\bf v}_{\rm D}^2/c^2}}\right).
\end{equation}
Taylor expanding the distribution function in powers of ${\bf v}_{\rm D}$ gives
\begin{equation}
{\bf J}_h=2\int \frac{d^dk}{(2\pi\hbar)^d}\,c^2{\bf k}
\left[f_0(\varepsilon_{\bf k})+\frac{\beta e^{\beta\varepsilon_{\bf
k}}}{(e^{\beta\varepsilon_{\bf k}}-1)^2}{\bf k}.{\bf v}_{\rm
D}+{\mathcal O}({\bf v}_{\rm D}^2)\right],
\end{equation}
where the first term vanishes upon integration. In the presence of an
electric field, $E_x$, and a magnetic field, $B_z\equiv B$, the drift
velocity ${\bf v}_{\rm D}$ is in the negative y-direction. As such, to
linear order in the electric field
\begin{equation}
J_h^y=-2\int \frac{d^dk}{(2\pi\hbar)^d}\,c^2k_y^2\frac{\beta
e^{\beta\varepsilon_{\bf k}}}{(e^{\beta\varepsilon_{\bf
k}}-1)^2}\frac{E_x}{B_z}.
\end{equation}
It follows from the defining relations (\ref{tcoeffs}) that
\begin{equation}
\tilde{\alpha}_{yx}=-\frac{2}{d}\int \frac{d^d k}{(2\pi\hbar)^d}\,
c^2k^2 \frac{\beta e^{\beta \varepsilon_{\bf
k}}}{(e^{\beta\varepsilon_{\bf k}}-1)^2} \frac{1}{B_z}.\end{equation}
Equivalently, using the Onsager relations\cite{Abrikosov:Fund}
$\tilde\alpha_{yx}({\bf B})=T\alpha_{yx}({\bf B})$ and
$\alpha_{xy}({\bf B})=\alpha_{yx}(-{\bf B})$
\begin{equation}
\alpha_{xy}=\frac{2c^2}{d B T}\int \frac{d^dk}{(2\pi\hbar)^d}\, 
k^2\left(-\frac{\partial f_0}{\partial\varepsilon_{\bf k}}\right).
\label{loralpha}
\end{equation}
This coincides with our previous result (\ref{enalpha}) obtained by
entropy drift arguments. This supports the validity of our simple
result, that at the strongly fluctuating SF-MI transition,
$\alpha_{xy}=2{\mathcal S}/B$, where ${\mathcal S}$ is the total
entropy density pertaining to each carrier type.\cite{Bhaseen:magneto}

\subsubsection{Linear Response}
\label{Sect:LR}

In order to go beyond our existing results, especially with a view to
thermal gradients in section \ref{Sect:TGM}, it is useful to perform a
systematic linear analysis of the QBE. In the presence of electric and
magnetic fields the Boltzmann equation reads
\begin{equation}
\frac{{\partial f}_{\pm}}{\partial t}\pm Q
\left({{\bf E}+{\bf v}_{\bf k}\times{\bf B}}\right).
\frac{\partial f_\pm}{\partial{\bf k}}={\rm I}_\pm[f_+,f_-].
\label{QBElin}
\end{equation}
To linear order in the electric field we may parameterize the
departure from equilibrium
\begin{equation}
f_\pm({\bf k})=f_0(\varepsilon_k)\pm Q{\bf k}.{\bf E}\,\psi(k)+{\bf k}.({\bf E}
\times{\bf B})\,\psi_\perp(k),
\label{ebexp}
\end{equation}
where for simplicity we focus on static solutions. Here, $\psi(k)$,
represents the longitudinal shift due to the applied electric field,
and $\psi_\perp(k)$ represents the transverse shift due to the
combined electric and magnetic field.  Substituting this expansion
into (\ref{QBElin}) and retaining only terms up ${\mathcal O}({\bf
E})$
\begin{equation}
\pm Q{\bf v}_{\bf k}.\left[{\bf E}\left(\frac{\partial
f_0}{\partial\varepsilon_k}\right)\mp Q({\bf E}\times{\bf
B})\,\psi(k)+{\bf E}|{\bf B}|^2\psi_\perp(k)\right]={\rm L}_\pm[\,\pm
Q{\bf k}.{\bf E}\,\psi\,]+{\rm L}_\pm^\prime[\,{\bf k}.({\bf
E}\times{\bf B})\,\psi_\perp \,],
\label{lineb}
\end{equation}
where the terms on the right hand side correspond to the distinct
linearizations of the collision term arising from the longitudinal and
transverse shifts of the distribution function --- see Appendix
\ref{App:Poly}. These are straightforward, but rather tedious to
derive, and involve momentum space integrals over the remaining
products of Bose distribution functions. As noted by Damle and
Sachdev,\cite{Damle:Nonzero,Damle:Thesis} these integrals may be
evaluated in terms of polylogarithm functions. Since the explicit form
of these linearizations was not included in any of the original
publications,\cite{Damle:Nonzero,Damle:Thesis,Bhaseen:magneto} we
provide them in Appendix \ref{App:Poly}. Although the details of the
collision term are certainly important for a quantitative numerical
implementation of the epsilon expansion, the physical results are
primarily determined by the robust symmetry properties of these
linearizations. In the case at hand, to lowest order in the epsilon
expansion the collision terms may be neglected in accordance with our
previous findings and we obtain
\begin{equation}
\psi(k)=0,\quad \psi_\perp(k)=\frac{1}{|{\bf B}|^2}
\left(-\frac{\partial f_0}{\partial\varepsilon_k}\right).
\label{psiperpeb}
\end{equation}
The vanishing of the longitudinal shift is consistent with the
vanishing of the DC electrical conductivity, $\sigma_{xx}(B)$, in the
drift regime.  The heat current is readily found from the expansion
(\ref{ebexp})
\begin{equation}
{\bf J}_h=2\int \frac{d^dk}{(2\pi\hbar)^d}\,
\varepsilon_k{\bf v}_k\,{\bf k}.({\bf E}\times{\bf B})\,\psi_\perp(k).
\end{equation}
As such, to lowest order in the epsilon expansion
\begin{equation}
{\bf J}_h=2c^2\int \frac{d^dk}{(2\pi\hbar)^d}\,{\bf k}\,({\bf k}.{\bf v}_{\rm D})
\left(-\frac{\partial f_0}{\partial\varepsilon_k}\right).
\label{driftjh}
\end{equation}
It is readily seen that this is consistent with our original transport
coefficient (\ref{enalpha}). Indeed, our lowest order linear response
solution (\ref{psiperpeb}), yields
\begin{equation}
f_\pm({\bf k})=f_0(\varepsilon_{\bf k})-{\bf k}.{\bf v}_{\rm D}
\left(\frac{\partial f_0}{\partial\varepsilon_k}\right).
\label{boostlin}
\end{equation}
This coincides with the linearization of our exact boost distribution
(\ref{boostdist}).

\subsubsection{Numerical Evaluation of $\alpha_{xy}$}
\label{Sect:Num}
Having presented compelling evidence for the general form of the
thermoelectric tensor, let us finally obtain the explicit numerical
result within the epsilon expansion.  Performing the angular integrals
yields
\begin{equation}
\alpha_{xy}=\frac{2c^2S_d}{dBT(2\pi\hbar)^d}\int_0^\infty dk\,k^{d+1}
\frac{\beta e^{\beta\varepsilon_{\bf k}}}{(e^{\beta\varepsilon_{\bf k}}-1)^2},
\end{equation}
where $S_d=2\pi^{d/2}/\Gamma(d/2)$ is the surface area of a unit
hypersphere in $d$-dimensions. Introducing rescaled variables ${\bar
k}\equiv ck/k_BT$ and $\tilde m\equiv mc^2/k_BT$ we find
\begin{equation}
\alpha_{xy}=\frac{k_B}{B}\left(\frac{k_BT}{\hbar c}\right)^d
\left[\frac{2S_d}{d(2\pi)^d}\int_0^\infty d{\bar k}\, {\bar k}^{d+1}
\frac{e^{-\sqrt{{\bar k}^2+{\tilde m}^2}}}{(1-e^{-\sqrt{{\bar k}^2+{\tilde m}^2}})^2}
\right].
\label{alphaexp}
\end{equation}
Within the epsilon expansion, $d=3-\epsilon$, and the mass parameter
entering equation (\ref{alphaexp}) is proportional to $\sqrt
{\epsilon}$ as indicated in (\ref{couplings}). To leading order we may
thus evaluate the dimensionless numerical prefactor in {\em three
dimensions} so as to obtain
\begin{equation}
\alpha_{xy}=\frac{2k_B}{B}\left(\frac{k_BT}{\hbar c}\right)^{3-\epsilon}
\left[\frac{1}{6\pi^2}\int_0^\infty d{\bar k}\, 
\frac{{\bar k}^4e^{-{\bar k}}}{(1-e^{-{\bar k}})^2}\right].
\end{equation}
The term in square brackets is just our massless entropy prefactor,
${\mathcal C}_3=2\pi^2/45$, and we thus find
\begin{equation}
\alpha_{xy}=\frac{4\pi^2}{45}\,\frac{k_B}{B}
\left(\frac{k_BT}{\hbar c}\right)^{3-\epsilon}.
\label{axygend}
\end{equation}
In particular, in $d=2$, one obtains
\begin{equation}
\alpha_{xy}=\alpha_0 \left(\frac{T^2}{B}\right).
\label{axyd2}
\end{equation}
where $\alpha_0 \approx 0.88\, k_B^3(\hbar c)^{-2}$. It is readily
seen that both answers (\ref{axygend}) and (\ref{axyd2}) are
consistent with the general scaling arguments presented in section
\ref{scaling}. Noting the inverse magnetic field dependence inherited
from the drift velocity (\ref{vdrift}), it follows from equation
(\ref{heatscal}) that the heat current scales as
\begin{equation}
J_h \sim T^{2+(d-1)/z}\frac{E}{T^{1+1/z}}\frac{T^{2/z}}{B}=
T\left(\frac{T^{d/z}}{B}\right)E={\tilde \alpha}E,
\end{equation}
in response to applied electric and magnetic fields. Our results are
therefore consistent with the scaling behavior
\begin{equation}
\alpha_{xy}\sim \frac{T^{d/z}}{B},
\end{equation}
when specialized to the Lorentz invariant case with $z=1$. This
completes our initial survey of the drift regime in crossed electric
and magnetic fields.

\subsection{Collision Dominated Regime: $|{\bf E}|>c|{\bf B}|$}
\label{Sect:CDR}

Having discussed the drift regime in considerable detail, let us very
briefly comment on our expectations when $E>cB$. As we discussed in
section \ref{sect:crossed}, a single particle is continually
accelerated by the electric field. Since the collisions in the QBE
(\ref{QBE}) conserve energy and momentum, there is no way to relax a
transverse heat current, and so we expect $\alpha_{xy}$ to diverge at
the clean fixed point in the absence of impurities. On the other hand,
the regime $E>cB$ is continuously connected to that studied by Damle
and Sachdev,\cite{Damle:Nonzero,Damle:Thesis} with $E\neq 0$ and
$B=0$. We thus expect the possibility of a finite electrical
conductivity, $\sigma_{xx}(B)$. It would be interesting to check these
expectations by including impurity scattering in the framework of the
QBE, although we do not pursue this here.

\section{Temperature Gradient and Magnetic Field} 
\label{Sect:TGM}

Having discussed the heat current which flows in response to an
electric field, we now turn our attention to the complementary problem
of the electric current which flows in response to a temperature
gradient. This is important in order to verify the Onsager reciprocity
relations,\cite{Abrikosov:Fund,Onsager:R1,Onsager:R2} which we have
used at several points to recast
$\tilde\alpha_{xy}=T\alpha_{xy}$. Although we no longer have the
luxury of Lorentz invariance arguments, we will again argue in favor
if two distinct regimes of behavior. We will begin in section
\ref{TDrift} with a discussion of the drift regime. In section
\ref{Sect:Moments} we will recover our previous results for
$\alpha_{xy}$, by taking an appropriate moment of the QBE in a thermal 
gradient. In
section \ref{Sect:LinT} we will further rederive this from a
linearization of the collision term. Both of these approaches indicate
the presence of a finite thermal conductivity, $\bar\kappa_{xx}(B)$,
and we will examine this in more detail in section \ref{Sect:TC}

\subsection{Drift Regime: $\nabla T\lesssim B$}
\label{TDrift}
\subsubsection{Transverse electrical current in response to a temperature gradient}
\label{Sect:Moments}
Although we have obtained $\alpha_{xy}$ by several different routes in
crossed electric and magnetic fields, it is prudent to consider the
computation in a thermal gradient. The celebrated Onsager
relations\cite{Abrikosov:Fund,Onsager:R1,Onsager:R2} tell us that we
ought to be able to compute $\alpha_{xy}$ by looking at the transverse
electrical current which flows in response to a temperature
gradient. Although this Onsager relation is expected to be true, it
ought to be verified by direct calculation. An additional motivation to 
examine this complementary approach is
that it will also pave the way to an analysis of the thermal
conductivity, $\bar\kappa(B)$. Let us therefore drop the electric
field from the Boltzmann equation (\ref{QBE}) and introduce a
temperature gradient. This is easily achieved by allowing the
temperature variable to be a function of
position.\cite{Ziman:Principles} The generic form of the Boltzmann
equation with ${\bf E}=0$ reads
\begin{equation}
\frac{\partial f_\pm}{\partial t}+{\bf v}_{\bf k}.\frac{\partial
f_\pm}{\partial{\bf x}}\pm Q({\bf v}_{\bf k}\times{\bf
B}).\frac{\partial f_\pm}{\partial {\bf k}}={\rm I}_\pm[f_+,f_-].
\end{equation}
In the absence of any material inhomogeneity we may assume that any
spatial variation is due to the imposed temperature gradient:
\begin{equation}
\frac{\partial f_\pm}{\partial {\bf x}}=\nabla_{\bf
x}T\left(\frac{\partial f_\pm}{\partial T}\right).
\end{equation}
Within linear response in $\nabla T$, we may replace $f_\pm$ by $f_0$
in the spatial gradient term. In this way we may write
\begin{equation}
\frac{\partial f_\pm}{\partial t}+{\bf v}_{\bf k}.\nabla_{\bf x}T
\left(-\frac{\varepsilon_k}{T}\frac{\partial f_0}{\partial
\varepsilon_k}\right)\pm Q({\bf v}_{\bf k}\times{\bf
B}).\frac{\partial f_\pm}{\partial {\bf k}}={\rm I}_\pm[f_+,f_-].
\label{betemo}
\end{equation}
A characteristic feature of the (single particle) drift regime is that
there is no net energy gain during each revolution; the energy gained
as a particle is accelerated under the electric field is lost on the
reverse journey.\cite{Jackson:Classical} This suggests that we ought
to look at the time variation of the total energy current within
Boltzmann theory. We may do so directly from equation (\ref{betemo})
by multiplying both sides by $\varepsilon_k{\bf v}_k$ and integrating
over all momenta. For relativistic particles this is aided by the fact
that ${\varepsilon}_k{\bf v}_k=c^2{\bf k}$ is proportional to the
momentum:
\begin{equation}
\frac{\partial {\bf J}_E^\pm}{\partial t}+c^4\int
\frac{d^dk}{(2\pi\hbar)^d}\,{\bf k}({\bf k}.{\bf
U})\left(\frac{\partial f_0}{\partial\varepsilon_k}\right)\pm Qc^2\int
\frac{d^dk}{(2\pi\hbar)^d}\,{\bf k}({\bf v}_k\times{\bf
B}).\frac{\partial f_\pm}{\partial{\bf k}}=0,
\end{equation}
where we define ${\bf U}\equiv (-\nabla T)/T$, and ${\bf J}_E^\pm$ are
the energy currents pertaining to particles and holes. In writing this
equation we have used the fact that the integral involving the
collision term vanishes; the momentum is an example of a so-called
summational invariant.\cite{Harris:BE} These quantities play an
important r\^ole in the hydrodynamic limit. Integrating the third term
by parts one obtains
\begin{equation}
\frac{\partial {\bf J}_E^\pm}{\partial t}+c^4\int
\frac{d^dk}{(2\pi\hbar)^d}\,{\bf k}({\bf k}.{\bf
U})\left(\frac{\partial f_0}{\partial\varepsilon_k}\right)\mp Qc^2\int
\frac{d^dk}{(2\pi\hbar)^d}\,({\bf v}_k\times{\bf B}) f_\pm =0,
\end{equation}
where we have used the fact that the terms involving derivatives of
${\bf v}_k$ vanish. Adding the particle and hole solutions yields
\begin{equation}
\frac{\partial {\bf J}_E}{\partial t}+2c^4\int
\frac{d^dk}{(2\pi\hbar)^d}\,{\bf k}({\bf k}.{\bf
U})\left(\frac{\partial f_0}{\partial\varepsilon_k}\right)-c^2{\bf
J}_e\times{\bf B}=0,
\end{equation}
where ${\bf J}_E$ and ${\bf J}_e$ are the total heat and electrical
currents defined by equations (\ref{bolel}) and (\ref{bolheat}). If we
impose the time independence of ${\bf J}_E$ we obtain
\begin{equation}
{\bf J}_e\times{\bf B}=2c^2\int \frac{d^dk}{(2\pi\hbar)^d}\,{\bf
k}({\bf k}.{\bf U})\left(\frac{\partial
f_0}{\partial\varepsilon_k}\right).
\end{equation}
In particular, if we apply a temperature gradient, $U_y$, in the
$y$-direction and a magnetic field $B_z\equiv B$ in the $z$-direction,
one obtains a transverse electrical current
\begin{equation}
-J_e^x B=2c^2\int
 \frac{d^dk}{(2\pi\hbar)^d}\,k_y^2\left(\frac{\partial f_0}{\partial
 \varepsilon_k}\right) U_y.
\end{equation}
From the defining relations (\ref{tcoeffs}) we thus obtain
\begin{equation}
\alpha_{xy}=\frac{2c^2}{dBT}\int\frac{d^dk}{(2\pi\hbar)^d}\,k^2\left(-\frac{\partial f_0}{\partial\varepsilon_k}\right).
\end{equation}
This is in agreement with our drift answer (\ref{enalpha}) and our
complementary calculation (\ref{loralpha}) based on Lorentz invariance
arguments in crossed ${\bf E}$ and ${\bf B}$ fields. By taking
appropriate moments of the Boltzmann equation\cite{Huang:Stat} we have
obtained the thermoelectric tensor without recourse to an explicit
solution. Moreover, we have recovered the correct Onsager symmetry
relation. In addition, the vanishing of ${\dot{\bf J}}_E$ indicates
that the thermal conductivity is {\em finite} in this regime. We shall
examine this further in section \ref{Sect:TC}. Before doing this we
first need to perform a linear response analysis in $\nabla T$.

\subsubsection{Explicit Construction of Linear Response in $\nabla T$}
\label{Sect:LinT}

In the presence of a temperature gradient and a magnetic field
\begin{equation}
\frac{\partial f_\pm}{\partial t}+c^2{\bf k}.{\bf
U}\left(\frac{\partial f_0}{\partial\varepsilon_k}\right)\pm Q({\bf
v}_k\times {\bf B}).\frac{\partial f_\pm}{\partial{\bf k}}={\rm
I}_\pm[f_+,f_-],
\label{beu}
\end{equation}
where ${\bf U}\equiv (-\nabla T)/T$. To linear order in the
temperature gradient we parameterize
\begin{equation}
f_\pm({\bf k})=f_0(\varepsilon_k)+{\bf k}.{\bf U}\,\psi(k)\pm Q{\bf
k}.({\bf U}\times {\bf B})\,\psi_\perp(k).
\label{betexp}
\end{equation}
Note that $\psi(k)$ and $\psi_\perp(k)$ are distinct from those
defined previously in equation (\ref{ebexp}). Substituting
(\ref{betexp}) into (\ref{beu}) and retaining only those terms of
${\mathcal O}({\bf U})$
\begin{equation}
c^2{\bf k}.\left[{\bf U}\left(\frac{\partial
f_0}{\partial\varepsilon_k}\right) \mp \frac{Q}{\varepsilon_k}({\bf
U}\times{\bf B})\,\psi(k)+\frac{Q^2{\bf U}|{\bf
B}|^2}{\varepsilon_k}\psi_\perp(k)\right]={\rm L}_\pm^\prime[\,{\bf
k}.{\bf U}\,\psi\,]+{\rm L}_\pm[\,\pm Q{\bf k}.({\bf U}\times{\bf
B})\,\psi_\perp\,],
\label{thermolin}
\end{equation}
where ${\rm L}$ and ${\rm L}^\prime$ are the distinct linearizations
of the collision term discussed in Appendix \ref{App:Poly}.  To lowest
order in the epsilon expansion we may drop the collision terms to
obtain
\begin{equation}
\psi(k)=0,\quad \psi_\perp(k)=\frac{\varepsilon_k}{Q^2|{\bf
B}|^2}\left(-\frac{\partial f_0}{\partial \varepsilon_k}\right).
\label{tempperp}
\end{equation}
The electrical current obtained from (\ref{betexp}) yields
\begin{equation}
{\bf J}_e=2Q^2\int \frac{d^dk}{(2\pi\hbar)^d}\,{\bf v}_k {\bf k}.({\bf
U}\times{\bf B})\psi_\perp(k).
\label{elcurtemp}
\end{equation}
Substituting (\ref{tempperp}) into (\ref{elcurtemp}) we find 
\begin{equation}
{\bf J}_e=2c^2\int \frac{d^dk}{(2\pi\hbar)^d}{\bf k}\,{\bf
k}.\left(\frac{{\bf U}\times{\bf B}}{|{\bf
B}|^2}\right)\left(-\frac{\partial f_0}{\partial\varepsilon_k}\right).
\end{equation}
This has a structure that is close to that of equation
(\ref{driftjh}), and once again this yields our previous expression
for $\alpha_{xy}$. However, in addition we will be able to go beyond
this result and examine the thermal conductivity,
$\bar\kappa_{xx}(B)$.

\subsubsection{Thermal Conductivity}
\label{Sect:TC}

\begin{figure}
\psfrag{k}{${\bar k}$} \psfrag{y}{$\bar k^4 G({\bar k})$}
\includegraphics[width=9cm,clip=true]{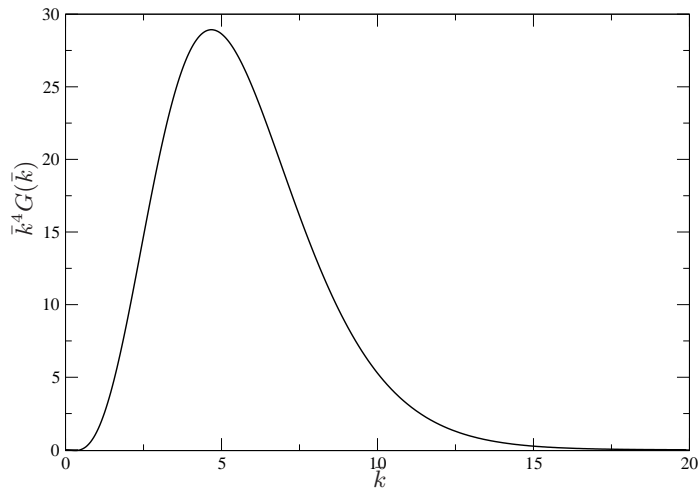}
\caption{Longitudinal scaling function, $G({\bar k})$, required for 
computation of the thermal conductivity in the drift regime. This figure 
was obtained by truncating the upper limit of integration at ${\bar k}_1=100$.}
\label{fig:kapscal}
\end{figure}

Thus far, we have used the QBE in a thermal gradient and a magnetic
field to verify our previous result for $\alpha_{xy}$, and the Onsager
relation. As we have seen, this is encoded in the lowest order
${\mathcal O}(\epsilon^0)$ solution given in equation
(\ref{tempperp}). However, self consistency of this non-trivial result
for $\psi_\perp(k)$, in turn induces a non-trivial {\em longitudinal}
shift (and a finite thermal conductivity) at the higher order of
${\mathcal O}(\epsilon^2)$.  As indicated by equation
(\ref{thermolin}), the ${\mathcal O}(\epsilon^0)$ solution for
$\psi_\perp(k)$ yields a non-trivial $\psi(k)$ at ${\mathcal
O}(\epsilon^2)$:
\begin{equation}
\mp Q{\bf v}_k.({\bf U}\times{\bf B})\,\psi(k)={\rm L}_\pm\left[\,\pm
Q{\bf k}.({\bf U}\times{\bf B})\,\psi_\perp(k)\right].
\label{kappapsi}
\end{equation}
The right hand side of this equation is structurally similar to the
linearization encountered by Damle and Sachdev\cite{Damle:Nonzero} in
the context of (zero magnetic field) electrical transport. The only
crucial difference is that the electric field has been replaced by
${\bf U}\times{\bf B}$, where we recall that ${\bf U}\equiv (-\nabla
T)/T$. Using their notation, equation (\ref{kappapsi}) may be written
\begin{equation}
\mp {\bf v}_k.{\bf C}\,\psi(k)=\pm{\bf k}.{\bf
C}\left\{-\epsilon^2\left(\frac{c}{\hbar}\right)\int_0^\infty
dk_1\left[\psi_\perp(k)\,{\rm F}_1(k,k_1)+\psi_\perp(k_1)\,{\rm
F}_2(k,k_1)\right]\right\},
\label{kaplinone}
\end{equation}
where we denote the combination $Q({\bf U}\times{\bf B})={\bf C}$ ---
see equation (\ref{compact}). Here, ${\rm F}_1(k,k_1)$ and ${\rm
F}_2(k,k_1)$ are non-trivial kernels which we provide in equations
(\ref{F1poly}), (\ref{F2apoly}) and (\ref{F2bpoly}) of Appendix
\ref{App:Poly}. In writing equation (\ref{kaplinone}) we have also
restored a factor of $c/\hbar$ which stems from requirement that the
collision term have dimensions of $s^{-1}$; the kernels are
dimensionless and the prefactor combines with the measure of
integration. That is to say,
\begin{equation}
\psi(k)=\epsilon^2\left(\frac{\varepsilon_k}{\hbar
c}\right)\int_0^\infty dk_1\left[\psi_\perp(k){\rm
F}_1(k,k_1)+\psi_\perp(k_1){\rm F}_2(k,k_1)\right].
\label{psieps}
\end{equation}
In view of the explicit $\epsilon^2$ dependence of equation
(\ref{psieps}) we should evaluate $\varepsilon_k$ and $\psi_\perp(k)$
in the {\em massless} limit.  As follows from equation
(\ref{tempperp})
\begin{equation}
\psi_\perp(k)\rightarrow \frac{1}{Q^2 B^2}\frac{{\bar k}e^{\bar
k}}{(e^{\bar k}-1)^2}; \quad {\bar k}\equiv \frac{ck}{k_BT}.
\label{mlesspsiperp}
\end{equation}
It is readily seen from Appendix \ref{App:Poly}, that the non-trivial
kernels ${\rm F}_1$ and ${\rm F}_2$, are in fact functions of the
dimensionless variables ${\bar k}$ and ${\bar k}_i$; note that we
worked in units where $\hbar=c=1$ there. Following Damle and
Sachdev,\cite{Damle:Nonzero} we denote
\begin{equation}
{\rm F}_1(k,k_1)\equiv {\Phi}_1({\bar k},{\bar k}_1),\quad {\rm
F}_2(k,k_1)\equiv {\Phi}_2({\bar k},{\bar k}_1).
\end{equation}
The longitudinal displacement may thus be written
\begin{equation}
\psi(k)=\frac{\epsilon^2}{\hbar}\left(\frac{k_BT}{Q B c}\right)^2G({\bar k}),
\label{glongi}
\end{equation}
where we introduce a {\em universal scaling function} of the
dimensionless variable $\bar k$
\begin{equation}
G(\bar k)\equiv {\bar k}\int_0^\infty d{\bar k}_1 \left[\frac{{\bar
k}e^{\bar k}}{(e^{\bar k}-1)^2}\, \Phi_1({\bar k},{\bar
k}_1)+\Phi_2({\bar k},{\bar k}_1)\, \frac{{\bar k}_1e^{{\bar
k}_1}}{(e^{{\bar k}_1}-1)^2}\right].
\label{gfunc}
\end{equation}
We plot an appropriate moment of this distribution in
Fig.~\ref{fig:kapscal}.  The heat current may be obtained from
(\ref{glongi}) by combining (\ref{betexp}) with the usual relation
(\ref{bolheat})
\begin{equation}
{\bf J}_h=2\int \frac{d^dk}{(2\pi\hbar)^d}\,\epsilon_k{\bf v}_k ({\bf
k}.{\bf U})\,\psi(k).
\end{equation}
It follows from the definitions (\ref{tcoeffs}) that the corresponding
thermal conductivity is given by
\begin{equation}
\bar\kappa_{xx}=\frac{2c^2}{dT}\int \frac{d^dk}{(2\pi\hbar)^d}\,k^2\,\psi(k).
\end{equation}
Performing the angular integrals gives
\begin{equation}
\bar\kappa_{xx}=\frac{2c^2}{dT}\frac{S_d}{(2\pi\hbar)^d}\int_0^\infty
dk\,k^{d+1}\,\psi(k), 
\label{kapparad}
\end{equation}
where $S_d=2\pi^{d/2}/\Gamma(d/2)$ is the surface area of a unit
hypersphere in $d$-dimensions. Substituting the result (\ref{glongi})
into (\ref{kapparad}) and rescaling the momentum integral yields
\begin{equation}
\bar\kappa_{xx}=\epsilon^2 \,k_B c \left(\frac{\hbar}{QB}\right)^2
\left(\frac{k_BT}{\hbar
c}\right)^{d+3}\left(\frac{2S_d}{d(2\pi)^d}\right)\int_0^\infty d{\bar
k}\, {\bar k}^{d+1}\,G({\bar k}).
\end{equation}
Equivalently 
\begin{equation}
\bar\kappa_{xx}=g\,\epsilon^2(k_Bc)\,{l}_B^4\,\lambda_T^{-(d+3)},
\label{kaplengths}
\end{equation}
where 
\begin{equation}
l_B\equiv \sqrt{\frac{\hbar}{QB}},\quad \lambda_T\equiv \frac{\hbar c}{k_BT},
\end{equation}
are the magnetic length, and a suitable thermal wavelength
respectively. The numerical coefficient is given by
\begin{equation}
g=\frac{2S_d}{d(2\pi)^d}\int_0^\infty d{\bar k}\,{\bar
  k}^{d+1}\,G({\bar k}).
\label{gdint}
\end{equation}
From equation (\ref{kaplengths}) we see that $\bar\kappa$ has
dimensions $Jk^{-1}m^{-(d-2)}s^{-1}$. This is consistent with the
defining relations (\ref{tcoeffs}). In addition, it is readily seen
that our expressions for the thermal conductivity are in agreement
with the scaling form (\ref{heatscal}) with dynamical exponent
$z=1$. To lowest order in the epsilon expansion, the integral
(\ref{gdint}) should be performed in $d=3$:
\begin{equation}
g=\frac{1}{3\pi^2}\int_0^\infty d{\bar k}\,{\bar k}^{4}\,G({\bar k}).
\end{equation}
This equation mirrors (3.33) of Damle and Sachdev.\cite{Damle:Nonzero}
The extra factor of momentum arises because we are considering {\em
heat} transport as opposed to electrical transport. In
Fig. \ref{fig:kapscal} we plot the universal scaling function ${\bar
k}^4\,G({\bar k})$.
We find by numerical integration that
\begin{equation}
g\approx 5.55.
\end{equation}
In particular, in $d=2$, we find
\begin{equation}
\bar\kappa_{xx}=\bar\kappa_0\left(\frac{T^5}{B^2}\right),
\end{equation}
where $\bar\kappa_0\equiv g k_B^6/(4e^2\hbar^3c^4)$. That is to say,
in stark contrast to the case where $B=0$, the thermal conductivity
does {\em not} diverge, but is {\em finite}.\cite{Bhaseen:magneto}
Moreover, the dependence on $\epsilon^2$ is {\em inversely} related to
that of the universal DC electrical conductivity
\begin{equation}
\sigma_Q=\frac{{\mathcal
N}_\sigma}{\epsilon^2}\left(\frac{4e^2}{h}\right),
\end{equation}
where ${\mathcal N}_\sigma\approx 1.037$. When specialized to
two-dimensions our result\cite{Bhaseen:magneto} (\ref{kaplengths}) may
be cast in the equivalent form\cite{Hartnoll:Nernst}
\begin{equation}
\bar\kappa_{xx}={\mathcal A}\left(\frac{{\mathcal S}_{\rm ch}^2 T}{B^2\sigma_Q}\right),
\label{PRLkappa}
\end{equation}
where, ${\mathcal S}_{\rm ch}= 2{\mathcal C}_3k_B\lambda_{\rm
T}^{-2}$, is the entropy density of a {\em charged} scalar field
calculated within the epsilon expansion, and the transformed
dimensionless prefactor is given by
\begin{equation}
{\mathcal A}\equiv \frac{g{\mathcal N}_\sigma}{8\pi{\mathcal C}_3^2}.
\end{equation}
With $g\approx 5.55$, ${\mathcal N}_\sigma\approx 1.037$, and 
${\mathcal C}_3=2\pi^2/45$ one obtains ${\mathcal A}\approx 1.19$.

\section{Hydrodynamic Limit and Interpolation}
\label{sect:hydro}
The arguments presented above are manifestly iterative in the epsilon
expansion and implicitly assume that $\epsilon^2\ll B$. It was
subsequently pointed out in Ref.~\onlinecite{Hartnoll:Nernst} using
relativistic hydrodynamic arguments {\em directly} in $d=2$, that
equation (\ref{PRLkappa}) holds as an {\em exact} relation in two
dimensions with ${\mathcal A}=1$, or $g_{{\rm MHD}}\approx
4.66$. Moreover, it was shown that all the physical response functions
are governed by combinations of thermodynamic variables and the single
parameter $\sigma_Q$. In order to make contact with the hydrodynamic
results of Ref. \onlinecite{Hartnoll:Nernst} we must study the ultra
low field limit $B\ll \epsilon^2$. Even though $\epsilon^2$ is a small
parameter, we must consider {\em both} the frequency {\em and} the
field scales much smaller than this in order to enter the hydrodynamic
regime. This is in conformity with the original zero field treatment
of Damle and Sachdev\cite{Damle:Nonzero} where the hydrodynamic
crossover occurs at $\omega\sim{\mathcal O}(\epsilon^2)$. The magnetic
field provides an additional cyclotron frequency and in the
hydrodynamic limit this is assumed to be less than the scattering
rate. As we shall discuss, the thermal conductivity in fact
interpolates between these two closely separated limits (${\mathcal
A}\approx 1.19$ and ${\mathcal A}=1$) as the ratio $B/\epsilon^2$ is
varied. This mirrors recent findings in strictly two-dimensional
graphene where (aside from the important aspects of spatial
dimensionality and carrier statistics) the fine structure constant,
$\alpha$, plays a similar r\^ole to $\epsilon$.\cite{Muller:Graphene}
In the subsequent discussion we derive the exact form of the
hydrodynamic result analytically, using the QBE and the epsilon
expansion around $d=3$. We will also discuss the interpolation as the
magnetic field is varied.

\begin{figure}
\psfrag{g}{$g$}
\psfrag{r}{$r$}
\includegraphics[clip=true,width=10cm]{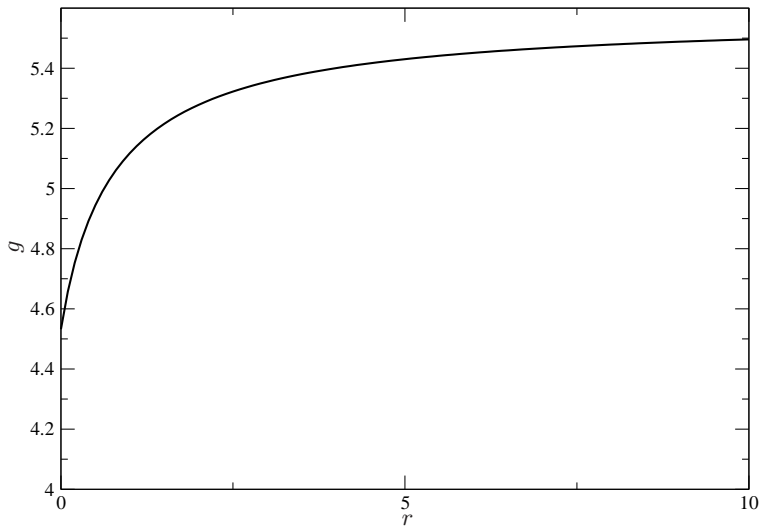}
\caption{Evolution of the dimensionless thermal conductivity
prefactor, $g$, as the ratio, $r\propto B^2/\epsilon^4$,
varies. Within the accuracy of our three dimensional Monte Carlo
integrations (approximately $3\%$) the result interpolates between the
value reported in our previous Letter, \cite{Bhaseen:magneto}
$g_\infty\approx 5.55$, and the magnetohydrodynamic
value,\cite{Hartnoll:Nernst} $g_0=8\pi(2\pi^2/45)^2/(1.037)\approx
4.66$. The latter corresponds to the exact relation,
$\bar\kappa_{xx}(B)=T{\mathcal S}_{\rm ch}^2/B^2\sigma_Q$, for $B\ll
\epsilon^2$.}
\label{Fig:gfactorevol}
\end{figure}

Returning to our original equation (\ref{thermolin}), we wish to solve
this coupled problem for the longitudinal and transverse distribution
functions in more generality. It is convenient to decompose this
relation in to two distinct equations which are respectively even and
odd under reversal of the magnetic field:
\begin{equation}
{\bf v}_k.{\bf U}\left[\varepsilon_k \frac{\partial f_0}{\partial\varepsilon_k}
+Q^2B^2\,\psi_\perp(k)\right]
={\rm L}_\pm^\prime[{\bf k}.{\bf U}\,\psi(k)],
\label{even}
\end{equation}
and
\begin{equation}
\mp Q{\bf v}_k.({\bf U}\times{\bf B})\,\psi(k) ={\rm L}_\pm [\pm Q{\bf
k}.({\bf U}\times{\bf B})\,\psi_\perp(k)].
\label{odd}
\end{equation}
Direct elimination will yield equations governing the longitudinal and
transverse response. In order to expose this more clearly, we use the
expressions (\ref{compact}) and (\ref{compactprime}) for the distinct
longitudinal and transverse linearizations of the collision term:
\begin{equation}
{\bf v}_k.{\bf U}\left[\varepsilon_k \frac{\partial f_0}{\partial\varepsilon_k}
+Q^2B^2\,\psi_\perp(k)\right]  = {\bf k}.{\bf
  U}\left\{-\epsilon^2\left(\frac{c}{\hbar}\right)\int_0^\infty dk_1[{\rm
    F}_1^\prime(k,k_1)\psi(k)+{\rm F}_2^\prime(k,k_1)\psi(k_1)]\right\},
\end{equation}
and 
\begin{equation}
\mp {\bf v}_k.{\bf C}\,\psi(k)=\pm {\bf k}.{\bf C}
\left\{-\epsilon^2\left(\frac{c}{\hbar}\right)\int_0^\infty dk_1 \left[{\rm
    F}_1(k,k_1)\psi_\perp(k)+{\rm F}_2(k,k_1)\psi_\perp(k_1)\right]\right\},
\end{equation}
where ${\bf C}\equiv Q({\bf U}\times{\bf B})$, and we have restored
the factors of $\hbar$ and $c$ in the collision terms. The factors of
${\bf k}.{\bf U}$ and ${\bf k}.{\bf C}$ are readily cancelled leaving
coupled integral equations for the distribution functions. For
pedagogical purposes it is convenient to discretize these equations
and write them in the simpler matrix form
\begin{equation}
\begin{aligned}
-\varepsilon_k \frac{\partial f_0}{\partial
 \varepsilon_k}-Q^2B^2\psi_{\perp,k} & =\left(\frac{\epsilon^2}{\hbar
 c}\right)\varepsilon_k {\rm M}_{k,k_1}^\prime\psi_{k_1},\\ \psi_k &
 =\left(\frac{\epsilon^2}{\hbar c}\right)\varepsilon_k{\rm M}_{k,k_1}
 \psi_{\perp,k_1},
\end{aligned}
\label{discoup}
\end{equation}
where we adopt the useful shorthand
\begin{equation}
\begin{aligned}
{\rm M}_{k,k_1}\psi_{\perp,k_1} & \equiv \int_0^\infty dk_1 \left[{\rm
F}_1(k,k_1)\psi_\perp(k)+{\rm F}_2(k,k_1)\psi_\perp(k_1)\right],\\
{\rm M}_{k,k_1}^\prime\psi_{k_1} & \equiv \int_0^\infty dk_1
\left[{\rm F}_1^\prime(k,k_1)\psi(k)+{\rm
F}_2^\prime(k,k_1)\psi(k_1)\right].
\label{Mdefs}
\end{aligned}
\end{equation}
Rearranging the first of equations (\ref{discoup}) for $\psi_\perp$
and substituting in to the second immediately yields a Fredholm
integral equation of the second kind for the longitudinal distribution
function
\begin{equation}
\psi_k=\psi_k^\infty-\left(\frac{\epsilon^2}{QB\hbar
c}\right)^2\varepsilon_k {\rm M}_{k,k_1}\varepsilon_{k_1}{\rm
M}^\prime_{k_1,k_2}\psi_{k_2},
\label{genpsieq}
\end{equation}
where 
\begin{equation}
\psi_k^\infty \equiv \left(\frac{\epsilon^2}{Q^2B^2 \hbar
c}\right)\varepsilon_k{\rm
M}_{k,k_1}\varepsilon_{k_1}\left(-\frac{\partial
f_0}{\partial\varepsilon_{k_1}}\right).
\label{psikinf}
\end{equation}
In the massless limit, we see that $\psi_\infty$ is nothing but our
previous solution obtained from equations (\ref{psieps}) and
(\ref{mlesspsiperp}). It is recovered from the general integral
equation (\ref{genpsieq}) in the limit $B\gg \epsilon^2$. Employing
dimensionless variables, $\bar k\equiv ck/k_BT$, and noting the
rescaling of the integration measures in (\ref{Mdefs}), the solutions
of equation (\ref{genpsieq}) are governed by the dimensionless
prefactor of the second term which we may denote as
\begin{equation}
r^{-1}\equiv {\epsilon^4}\left(\frac{l_B}{\lambda_T}\right)^4,
\end{equation}
where $l_B\equiv\sqrt{\hbar/QB}$ is the magnetic length, and
$\lambda_T\equiv \hbar c/k_BT$ is the thermal
wavelength. Equivalently,
\begin{equation}
r \sim (\omega_c^{\rm typ}\tau_{in})^2,
\end{equation}
where $\tau_{in}^{-1}\sim \epsilon^2 k_BT/\hbar$ is the inelastic
scattering rate due to the collisions at the clean fixed point, and
$\omega_c^{\rm typ}\sim QB/(k_BT/c^2)$ is the typical cyclotron rate
of a thermal carrier.\cite{Muller:Graphene} In the limit,
$r\rightarrow 0$, the particles experience a large number of
collisions per typical period of revolution in the magnetic
field. This corresponds to the hydrodynamic limit studied in Refs
\onlinecite{Hartnoll:Nernst,Muller:Graphene}. In this respect, the
magnetohydrodynamic parameter, $r$, plays a similar r\^ole to the
Knudsen parameter in the development of Chapman--Enskog
theory.\cite{Harris:BE} Deep in this hydrodynamic regime, $B\ll
\epsilon^2$, the longitudinal distribution function satisfies the {\em
homogeneous} equation
\begin{equation}
{\rm M}_{k,k_1}\varepsilon_{k_1}{\rm M}^\prime_{k_1,k_2}\psi_{k_2}=0,
\end{equation}
as follows directly from equation (\ref{genpsieq}). In this limit, the
distribution function projects on to a zero mode of the transverse
operator ${\rm M}^\prime$. From our previous discussion of the
transverse response (\ref{boostlin}) it is readily seen that
\begin{equation}
f_\pm(k)=f_0(k)+{\mathcal N}{\bf k}.{\bf U}\left(-\frac{\partial
f_0}{\partial\varepsilon_k}\right)+{\rm transverse}
\end{equation}
is a zero mode of this collision operator. Moreover, this form would
also emerge in a relaxation time approximation where ${\mathcal
N}\equiv \tau_\gamma c^2$ and $\tau_\gamma$ is a suitable time
scale. With the parameterization (\ref{betexp}) we therefore consider
\begin{equation}
\psi_k={\mathcal N}\left(-\frac{\partial f_0}{\partial \varepsilon_k}\right),
\label{hydropsi}
\end{equation}
where ${\mathcal N}$ is a dimensionfull parameter to be determined.
This normalization is fixed by the original {\em inhomogeneous}
equation (\ref{genpsieq}), and may be found by taking moments and
extrapolating to the appropriate hydrodynamic limit. Placing the last
term of equation (\ref{genpsieq}) on the left hand side, and inverting
the matrix operation in the definition of $\psi_k^\infty$, equation
(\ref{genpsieq}) may be rewritten in the equivalent form
\begin{equation}
\left(\frac{Q^2B^2\hbar c}{\epsilon^2}\right) \left[({\rm
M}^{-1})_{k,k_1}\varepsilon_{k_1}^{-1}\psi_{k_1}+\left(\frac{\epsilon^2}{QB\hbar
c}\right)^2\varepsilon_k{\rm M}_{k,k_1}^\prime\psi_{k_1}\right]
=\varepsilon_k\left(-\frac{\partial f_0}{\partial
\varepsilon_k}\right).
\end{equation}
In order to make contact with the functional dependence
(\ref{PRLkappa}) observed in the complementary regime, it is
convenient to multiply this equation by $\varepsilon_k$ and integrate
over all momenta. Upon sending $\psi_k$ to the hydrodynamic form
(\ref{hydropsi}), and working in the massless limit where
$\varepsilon_k=ck$, one then obtains the normalization condition
\begin{equation}
{\mathcal N} Q^2B^2\hbar c \int \frac{d^dk}{(2\pi\hbar)^d}\,
k\,\epsilon^{-2}({\rm M}^{-1})_{k,k_1} k_1^{-1}\left(-\frac{\partial
f_0}{\partial\varepsilon_{k_1}}\right)=c^2 \int
\frac{d^dk}{(2\pi\hbar)^d}\,k^2 \left(-\frac{\partial
f_0}{\partial\varepsilon_k}\right),
\label{normeq}
\end{equation}
where we exploit the fact that $\psi_k$ is a zero mode of ${\rm
M}^\prime$. Using equation (\ref{enalpha}) we see that the right hand
side of this equation is related to the entropy density. Moreover,
using equation (3.28) of Damle and Sachdev,\cite{Damle:Nonzero} we see
that the left hand side involves the DC conductivity in the {\em
absence} of a magnetic field. Recalling the main steps, we
parameterize, $f_\pm=f_0\pm Q{\bf k}.{\bf E}\,\psi_{\rm DS}(k)$, and
substitute into the QBE equation (\ref{QBElin}) with ${\bf B}=0$. To
linear order in the electric field one obtains
\begin{equation}
\frac{c}{k}\frac{\partial f_0}{\partial\varepsilon_k}=
-\epsilon^2\left(\frac{c}{\hbar}\right){\rm M}_{k,k_1}\psi_{{\rm
    DS}}(k_1).
\end{equation}
With this identification equation (\ref{normeq}) may be recast in the
form
\begin{equation}
{\mathcal N} Q^2B^2 c \int \frac{d^dk}{(2\pi\hbar)^d}\,k\,\psi_{\rm
DS}(k)=dT\frac{{\mathcal S}_{\rm ch}}{2},
\end{equation}
where ${\mathcal S}_{\rm ch}$ is the entropy density of a charged
scalar field.  This may be rearranged to read
\begin{equation}
{\mathcal N}B^2\left[\frac{2Q^2c}{d}\int
\frac{d^dk}{(2\pi\hbar)^d}\,k\,\psi_{\rm DS}(k)\right]=T{\mathcal
S}_{\rm ch}.
\end{equation}
Since the electric current is given by
\begin{equation}
{\bf J}_e=2Q^2\int \frac{d^dk}{(2\pi\hbar)^d}{\bf v}_k{\bf k}.{\bf
E}\,\psi_{\rm DS}(k),
\end{equation}
the quantitity in square brackets is the conductivity. That is to say
\begin{equation}
{\mathcal N}=\frac{T{\mathcal S}_{\rm ch}}{B^2\sigma_{Q}},
\label{psinorm}
\end{equation}
where $\sigma_Q$ is the universal and non-trivial value of the DC
electrical conductivity computed within the epsilon expansion in the
{\em absence} of a magnetic field.\cite{Damle:Nonzero} From this
normalization, ${\mathcal N}\equiv \tau_\gamma c^2$, we may extract
the characteristic damping time scale, $\tau_\gamma$, of the
collective cyclotron mode discussed by Hartnoll {\em et
al}.\cite{Hartnoll:Nernst} Returning to our distribution function
(\ref{hydropsi}), the heat current is given by
\begin{equation}
{\bf J}_h=2{\mathcal N}\int \frac{d^dk}{(2\pi\hbar)^d}\,\varepsilon_k
{\bf v}_k{\bf k}.{\bf U}\left(-\frac{\partial
f_0}{\partial\varepsilon_k}\right).
\end{equation}
The corresponding thermal conductivity reads
\begin{equation}
{\bar \kappa}_{xx}=\frac{2{\mathcal
N}c^2}{dT}\int\frac{d^dk}{(2\pi\hbar)^d}\,k^2\left(-\frac{\partial
f_0}{\partial\varepsilon_k}\right)\equiv {\mathcal N}{\mathcal S}_{\rm
ch},
\label{hydrotherm}
\end{equation}
where we employ equation (\ref{enalpha}) again. Combining equations
(\ref{psinorm}) and (\ref{hydrotherm}) one obtains the relation
\begin{equation}
\bar\kappa_{xx}=\frac{T{\mathcal S}_{\rm ch}^2}{B^2\sigma_Q},
\label{duality}
\end{equation}
deep in the hydrodynamic limit where $B\ll \epsilon^2$. We see that
the functional dependence is the same in both limits, $B\gg
\epsilon^2$, and $B\ll \epsilon^2$, and only the dimensionless
prefactor is modified by a factor close to unity. As discussed in
Appendix \ref{App:Longcross}, numerical solution of the integral
equation (\ref{genpsieq}) yields the interpolation between these two
regimes --- see Fig.~\ref{Fig:gfactorevol}.
It is instructive to note that the hydrodynamic result
(\ref{duality}), which holds exactly in two dimensions, is accessible
within the framework of the epsilon expansion about $d=3$. The main
delicate points are that we should work in a regime where
$B\ll\epsilon^2$, and be careful to interpret the observables
$\bar\kappa_{xx}$, $\sigma_{Q}$, and ${\mathcal S}_{\rm ch}$ by means
of their respective epsilon expansions.
\section{Conclusions}
\label{sect:conc}
In this work we have examined the magnetothermoelectric response in
the vicinity of a quantum critical point. We investigate the
electrical and thermal transport and thermodynamics, and have
presented general scaling arguments valid for arbitrary dimension,
dynamical exponent and carrier statistics. These are supported by
explicit calculations at the particle-hole symmetric SF-MI transitions
of the Bose--Hubbard model. The presence of a magnetic field strongly
influences the physical response, and we demonstrate the existence of
a finite thermoelectric tensor, $\alpha_{xy}$, and a finite thermal
conductivity, $\bar\kappa_{xx}(B)$, even in the absence of
impurities. We relate these observations to a number of different
approaches, based on Lorentz invariance, the quantum Boltzmann
equation (QBE), and field theory considerations.  In accordance with
recent findings of M\"uller {\em et al} for
graphene,\cite{Muller:Graphene} the dimensionless prefactor of our
thermal conductivity is a smoothly varying function of
$\omega_c\tau_{\rm in}$. We derive an illuminating integral equation
to describe this evolution, which exemplifies the r\^ole of zero modes
in the hydrodynamic limit.\cite{Muller:Graphene} Our analytic and
numerical calculations smoothly interpolate between the result
presented in our previous Letter,\cite{Bhaseen:magneto} and the limit
of two-dimensional relativistic
magnetohydrodynamics.\cite{Hartnoll:Nernst} The recovery of the
relativistic hydrodynamic relations within the framework of the
epsilon expansion around three dimensions is quite
compelling. Although the epsilon expansion is well established for the
calculation of critical indices relating to thermodynamic quantities,
it is much less widely employed in transport situations. The present
body of results clearly demonstrate that we may address critical
fluctuations in transport coefficients by using such methods. We see
that the QBE approach not only has broad applicability but also
provides a physically intuitive way to incorporate both quantum and
thermal fluctuations in the hydrodynamic regime.  Moreover, the
results obtained are physically transparent and provide a platform for
further studies.

\acknowledgments

We are extremely grateful to D. Basko, J.-S. Caux, J. Chalker,
K. Damle, F. Essler, C. Hooley, D. Huse, A. Lamacraft,
M. M\"uller, S. Sachdev, V. Oganesyan, B. Simons, and A. Tsvelik for
helpful discussions at various stages of this work. This work was
supported by NSF Grant No. DMR0213706, EPSRC Grants EP/E018130/1,
EP/D050952/1, EP/D036194/1, and The Royal Society. We are particularly
grateful to David Huse for insightful comments on the Lorentz
transformations. MJB would also like to thank BNL and St. Andrews for
hospitality during part of this work, the Rudolf Peierls Centre for
Theoretical Physics, and the Cavendish Laboratory for financial
support.

\appendix

\section{Collision Term and Polylogarithms}
\label{App:Poly}

In order to make progress with the Boltzmann equation it is useful to
linearize about an equilibrium Bose distribution. There are two
distinct linearizations depending on the context. We gather some
useful formulae below.

\subsection{$\delta f_\pm ({\bf k})=\pm {\bf k}.{\bf C}\,h(k)$}
In this linearization we consider departures from equilibrium of the
form $\delta f_\pm({\bf k})=\pm {\bf k}.{\bf C}\,h(k)$ where ${\bf C}$
is a constant vector, and $h(k)$ is a function of $|{\bf k}|$. This
situation arises in the longitudinal response to an electric
field,\cite{Damle:Nonzero} ${\bf C}=Q{\bf E}$, and the transverse
response to a temperature gradient, ${\bf C}=Q({\bf U}\times{\bf B})$.
We want to expand the collision term (\ref{col}) to linear order in
the applied field ${\bf C}$.  We may write
\begin{equation}
{\rm I}_\pm=-\frac{2u_0^2}{9}
\int {d\boldsymbol\mu}\,
\left({\mathcal F}_\pm^{\rm out}-{\mathcal F}_\pm^{\rm in}\right),
\label{appcol}
\end{equation}
where, in units where $\hbar=c=1$,
\begin{equation}
d{\boldsymbol\mu}\equiv \frac{1}{2\varepsilon_k}\left[\prod_{i=1}^3
\frac{d^dk_i}{(2\pi)^d}\frac{1}{2\varepsilon_{k_i}}\right](2\pi)^d\delta({\bf
k}+{\bf k}_1-{\bf k}_2-{\bf
k}_3)\,(2\pi)\delta(\varepsilon+\varepsilon_1-\varepsilon_2-\varepsilon_3)
\end{equation}
represents the remaining phase space measure in equation (\ref{col}).
It is readily verified that
\begin{eqnarray}
{\mathcal F}_\pm^{\rm out}-{\mathcal F}_\pm^{\rm in} & = & \pm 3{\bf
k}.{\bf C}\,h(k)\left(e^{\beta(\varepsilon_2+\varepsilon_3)}
-e^{\beta\varepsilon_1}\right)n(\varepsilon_1)
n(\varepsilon_2)n(\varepsilon_3)\nonumber \\ & & \pm {\bf k}_1.{\bf C}
\, h(k_1)\left(e^{\beta\varepsilon_k}
-e^{\beta(\varepsilon_2+\varepsilon_3)}\right)
n(\varepsilon_k)n(\varepsilon_2)n(\varepsilon_3) \nonumber \\ & & \pm
3{\bf k}_2.{\bf C}\,h(k_2)\left(e^{\beta\varepsilon_3}
-e^{\beta(\varepsilon_k+\varepsilon_1)}\right)
n(\varepsilon_k)n(\varepsilon_1)n(\varepsilon_3) \nonumber \\ & & \pm
{\bf k}_3.{\bf C}\,h(k_3)\left(e^{\beta(\varepsilon_k+\varepsilon_1)}
-e^{\beta\varepsilon_2}\right)
n(\varepsilon_k)n(\varepsilon_1)n(\varepsilon_2) +{\mathcal O}({\bf
C}^2),
\label{pmg}
\end{eqnarray}
where $n(\varepsilon)\equiv f_0(\varepsilon)$ is the Bose distribution
function, and we have used the identity
$1+n(\varepsilon)=e^{\beta\varepsilon}n(\varepsilon)$. Upon
substituting (\ref{pmg}) into (\ref{appcol}) it is convenient to
interchange ${\bf k}_3\leftrightarrow {\bf k}_2$ in the last
term. This transformation preserves the integration measure and the
linearized collision term becomes
\begin{equation}
{\rm I}_\pm\rightarrow{\rm L}_\pm=\pm\left(-\frac{2u_0^2}{9}\right)
\int {d{\boldsymbol\mu}}\,\left[{\bf k}.{\bf C}\,h(k)
{\rm T}_1(k_1,k_2,k_3)+{\bf k}_1.{\bf C}\,h(k_1){\rm T}_2(k,k_2,k_3)
+{\bf k}_2.{\bf C}\,h(k_2)\,{\rm T}_3(k,k_1,k_3)\right]
\end{equation}
where 
\begin{eqnarray}
{\rm T}_1(k_1,k_2,k_3) & \equiv &
3\left(e^{\beta({\varepsilon_2+\varepsilon_3)}}-e^{\beta\varepsilon_1}\right)n(\varepsilon_1)n(\varepsilon_2)n(\varepsilon_3),\\
{\rm T}_2(k,k_2,k_3) & \equiv &
\left(e^{\beta\varepsilon_k}-e^{\beta(\varepsilon_2+\varepsilon_3)}\right)n(\varepsilon_k)n(\varepsilon_2)n(\varepsilon_3),
\\ {\rm T}_3(k,k_1,k_3) & \equiv &
2\left(e^{\beta\varepsilon_3}-e^{\beta(\varepsilon_k+\varepsilon_1)}\right)n(\varepsilon_k)n(\varepsilon_1)n(\varepsilon_3).
\end{eqnarray}
As discussed by Damle and Sachdev,\cite{Damle:Nonzero} to leading
order in the epsilon expansion one may evaluate the necessary
integrals directly in $d=3$. In addition, one may consider the
massless limit where ${\varepsilon}_k={k}$, and we have set $c=1$. The
angular integrals and one of the radial integrals may be carried out
explicitly, by means of formulas (C1), (C3) and (C5) of
Ref. \onlinecite{Damle:Nonzero}; in their notations $d^3k_i \equiv
k_i^2 dk_i d\Omega_i$. Interchanging ${\bf k}_1\leftrightarrow {\bf
k}_3$ in the first term, and ${\bf k}_1\leftrightarrow {\bf k}_2$ in
the last, this procedure yields
\begin{eqnarray}
{\rm L}_\pm & = & \pm\left(-\frac{2u_0^2}{9}\right)\frac{{\bf k}.{\bf
C}}{(4\pi)^3}\int_0^\infty dk_1dk_2\,\left\{\frac{h(k)}{k^{2}}\,{\rm
T}_1(k_1+k_2-k,k_2,k_1)\,{\rm I}_1(k,k_1,k_2)\right. \nonumber \\ & &
\left.-\frac{h(k_1)}{3k^4}\left[\,{\rm T}_2(k,k_2,k+k_1-k_2)\,{\rm
I}_2(k,k_1,k_2)-{\rm T}_3(k,k_2,k+k_2-k_1){\rm
I}_3(k,k_1,k_2)\right]\right\},
\end{eqnarray}
where ${\rm I}_1$, ${\rm I}_2$ and ${\rm I}_3$ are (domain dependent)
polynomials given in (C2), (C4) and (C6) of
Ref. \onlinecite{Damle:Nonzero}. In this way one may write the
linearized collision term in the compact form\cite{Damle:Nonzero}
\begin{equation}
{\rm L}_\pm = \pm {\bf k}.{\bf C}\left\{-\epsilon^2\int_0^\infty
dk_1\,\left[h(k){\rm F}_1(k,k_1)+h(k_1){\rm
F}_2(k,k_1)\right]\right\},
\label{compact}
\end{equation}
where we use the fact that $u_0=(24/5)\pi^2\epsilon$, and we define
\begin{equation}
{\rm F}_1(k,k_1) = \frac{2\pi}{25k^2}\int_0^\infty dk_2\,{\rm
T}_1(k_1+k_2-k,k_2,k_1)\,{\rm I}_1(k,k_1,k_2),
\label{f1k}
\end{equation}
and 
\begin{eqnarray}
{\rm F}_2(k,k_1) & = & -\frac{2\pi}{75 k^4}\int_0^\infty
dk_2\,\left[{\rm T}_2(k,k_2,k+k_1-k_2)\,{\rm I}_2(k,k_1,k_2)
\right. \nonumber \\ & & \left. \hspace{3cm} -{\rm
T}_3(k,k_2,k+k_2-k_1){\rm I}_3(k,k_1,k_2)\right].
\label{f2k}
\end{eqnarray}
We again emphasize that the functions ${\rm T}_1$, ${\rm T}_2$ and
${\rm T}_3$ are evaluated in the {\em massless} limit.  More
explicitly
\begin{equation}
{\rm F}_1(k,k_1)=\frac{6\pi}{25}\frac{n(k_1)}{k^2n(k)}\int_0^\infty
dk_2\, n(k_2)\left[1+n(k_1+k_2-k)\right]{\rm I}_1(k,k_1,k_2).
\label{f1intkern}
\end{equation}
Likewise, if we denote
\begin{equation}
{\rm F}_2(k,k_1)\equiv {\rm F}_2^a(k,k_1)+{\rm F}_2^b(k,k_1)
\label{f2intkern}
\end{equation}
then 
\begin{eqnarray}
{\rm F}_2^a(k,k_1) & = & \frac{2\pi}{75}\frac{[1+n(k)]}{k^4n(k_1)}
\int_0^\infty dk_2 \,n(k_2)n(k+k_1-k_2)\,{\rm I}_2(k,k_1,k_2),
\label{f2aik}
\end{eqnarray}
\begin{eqnarray}
{\rm F}_2^b(k,k_1) & = &
-\frac{4\pi}{75}\frac{n(k)}{k^4n(k_1)}\int_0^\infty dk_2\,
n(k_2)[1+n(k+k_2-k_1)]\,{\rm I}_3(k,k_1,k_2).
\label{f2bik}
\end{eqnarray}
These expressions are in conformity with equations (C7) and (3.28) of
Ref. \onlinecite{Damle:Nonzero}. As noted by Damle and Sachdev, the
integrals (\ref{f1intkern}), (\ref{f2aik}) and (\ref{f2bik}) may be
evaluated exactly using polylogarithm functions. Although the method
was carefully explained, the explicit form of these kernels was not
stated in their original works.\cite{Damle:Nonzero,Damle:Thesis} The
expressions are quite lengthy, and are rather tedious to
derive. Nonetheless, they are valuable for numerical work. We may
write
\begin{equation}
{\rm
F}_1(k,k_1)=\frac{6\pi}{25}\frac{n(k_1)n(k-k_1)}{k^2\,n(k)}\left[\Theta(k-k_1)\mu_2(k,k_1)-\Theta(k_1-k)\mu_2(k_1,k)\right],
\label{F1poly}
\end{equation}
together with
\begin{equation}
{\rm
F}_2^a(k,k_1)=\frac{2\pi}{75}\frac{[1+n(k)]n(k+k_1)}{k^4\,n(k_1)}\,{\rm
L}_2^a(k,k_1),
\label{F2apoly}
\end{equation}
and
\begin{equation}
{\rm F}_2^b(k,k_1)
=\frac{4\pi}{75}\frac{n(k)n(k_1-k)}{k^4\,n(k_1)}\left[\Theta(k-k_1){\rm
L}_2^b(k,k_1)-\Theta(k_1-k){\rm L}_2^b(k_1,k)\right],
\label{F2bpoly}
\end{equation}
where $\Theta(k-k_1)$ is the step function. In writing these kernels
we have introduced
\begin{equation}
{\rm L}_2^a(k,k_1)=24\lambda^{-}_4(k,k_1)+12[\,k\,\eta_3(k,k_1)+k_1\eta_3(k_1,k)\,]-6kk_1\lambda^+_2(k,k_1),
\end{equation}
where
\begin{eqnarray}
\lambda^\pm_n(x,y) & \equiv & \beta^{-n}\left[{\rm Li}_n(e^{-\beta
x})+{\rm Li}_n(e^{-\beta y})\pm {\rm Li}_n(e^{-\beta(x+y)})\pm {\rm
Li}_n(1)\right], \nonumber \\ \eta_n(x,y) & \equiv &
\beta^{-n}\left[{\rm Li}_n(e^{-\beta x})-{\rm Li}_n(e^{-\beta y})-{\rm
Li}_n(e^{-\beta(x+y)})+{\rm Li}_n(1)\right]. \nonumber
\end{eqnarray}
Further,
\begin{equation}
{\rm L}_2^b(k,k_1)=-3\left[4\mu_4+2(k-k_1)\mu_3-kk_1\mu_2+4k_1\nu_3+2kk_1\nu_2\right],
\end{equation}
where
\begin{equation*}
\mu_n(x,y)\equiv \beta^{-n}\left[{\rm Li}_n(1)+{\rm Li}_n(e^{-\beta x})-{\rm Li}_n(e^{-\beta y})-{\rm Li}_n(e^{-\beta(x-y)})\right],
\end{equation*}
and
\begin{equation*}
\nu_n(x,y)\equiv \beta^{-n}\left[{\rm Li}_n(e^{-\beta x})-{\rm Li}_n(e^{-\beta y})\right].
\end{equation*}
Here ${\rm Li}_p(z)$ is the polylogarithm with series expansion
\begin{equation}
{\rm Li}_p(z)=\sum_{n=1}^\infty\frac{z^n}{n^p}.
\end{equation}
Note that we have used the fact that ${\rm L}_2^a(k,k_1)$ is {\em
symmetric} in order to eliminate the step functions from ${\rm
F}_2^a(k,k_1)$.  In writing these expressions we employ polylogarithms
whose arguments lie within the unit disc. Although tedious to check
analytically, the equality of these functions and the integral
representations (\ref{f1intkern}), (\ref{f2aik}) and (\ref{f2bik}) is
readily verified numerically.  It is worth noting that the kernels
${\rm F}_1(k,k_1)$ and ${\rm F}_2(k,k_1)$ possess singularities when
their arguments coincide.

\subsection{$\delta f_\pm ({\bf k})={\bf k}.{\bf C}\,h(k)$}

In this linearization we consider departures from equilibrium of the
form $\delta f_\pm({\bf k})={\bf k}.{\bf C}\,h(k)$ where ${\bf C}$ is
an arbitrary vector, and $h(k)$ is a function of $|{\bf k}|$. This
charge independent situation arises in the longitudinal response to a
temperature gradient, ${\bf C}={\bf U}\equiv(-\nabla T)/T$, and in the
transverse response to an electric field, ${\bf C}={\bf E}\times{\bf
B}$. In this case
\begin{eqnarray}
{\mathcal F}_\pm^{\rm out}-{\mathcal F}_\pm^{\rm in} & = & 3{\bf k}.{\bf C}\,
h(k)\left(e^{\beta(\varepsilon_2+\varepsilon_3)}-e^{\beta\varepsilon_1}\right)
n(\varepsilon_1)n(\varepsilon_2)n(\varepsilon_3) \nonumber \\
& + & 3{\bf k}_1.{\bf C}\,h(k_1)
\left(e^{\beta(\varepsilon_2+\varepsilon_3)}-e^{\beta\varepsilon_k}\right)
n(\varepsilon_k)n(\varepsilon_2)n(\varepsilon_3) \nonumber \\
& + & 3{\bf k}_2.{\bf C}\,h(k_2)
\left(e^{\beta\varepsilon_3}-e^{\beta(\varepsilon_k+\varepsilon_1)}\right)
n(\varepsilon_k)n(\varepsilon_1)n(\varepsilon_3) \nonumber \\
& + & 3{\bf k}_3.{\bf C}\,h(k_3)
\left(e^{\beta\varepsilon_2}-e^{\beta(\varepsilon_k+\varepsilon_1)}\right)
n(\varepsilon_k)n(\varepsilon_1)n(\varepsilon_2)+{\mathcal O}({\bf C}^2).
\end{eqnarray}
Substituting this into the collision term, and again making the interchange 
${\bf k}_3\leftrightarrow {\bf k}_2$ in the last term, we find
\begin{equation}
{\rm I}_\pm\leftrightarrow {\rm L}_\pm^{\prime}=\left(-\frac{2u_0^2}{9}\right)
\int d{\boldsymbol\mu}\,
\left[{\bf k}.{\bf C}\,h(k)\,{\rm T}_1^\prime(k_1,k_2,k_3)
+{\bf k}_1.{\bf C}\,h(k_1)\,{\rm T}_2^\prime(k,k_2,k_3)
+{\bf k}_2.{\bf C}\,h(k_2)\,{\rm T}_3^\prime(k,k_1,k_3)\right],
\end{equation}
where
\begin{eqnarray}
{\rm T}_1^\prime(k_1,k_2,k_3) & \equiv &
3\left(e^{\beta(\varepsilon_2+\varepsilon_3)}-e^{\beta\varepsilon_1}\right)
n(\varepsilon_1)n(\varepsilon_2)n(\varepsilon_3) \nonumber \\ {\rm
T}_2^\prime(k,k_2,k_3) & \equiv &
3\left(e^{\beta(\varepsilon_2+\varepsilon_3)}-e^{\beta\varepsilon_k}\right)
n(\varepsilon_k)n(\varepsilon_2)n(\varepsilon_3) \nonumber \\ {\rm
T}_3^\prime(k,k_1,k_3) & \equiv &
6\left(e^{\beta\varepsilon_3}-e^{\beta(\varepsilon_k+\varepsilon_1)}\right)
n(\varepsilon_k)n(\varepsilon_1)n(\varepsilon_3).
\end{eqnarray}
In particular, it is readily seen that ${\rm T}_1^\prime={\rm T}_1$,
${\rm T}_2^\prime=-3{\rm T}_2$ and ${\rm T}_3^\prime=3{\rm T}_3$. This
procedure therefore yields
\begin{eqnarray}
{\rm L}_\pm^\prime & = & \left(-\frac{2u_0^2}{9}\right)
\frac{{\bf k}.{\bf C}}{(4\pi)^3}
\int_0^\infty dk_1dk_2\left\{\frac{h(k)}{k^2}\,{\rm T}_1(k_1+k_2-k,k_2,k_1)\,{\rm I}_1(k,k_1,k_2)\right. \nonumber \\
& & \left. +\frac{h(k_1)}{k^4}\left[{\rm T}_2(k,k_2,k+k_1-k_2)\,{\rm I}_2(k,k_1,k_2)
+{\rm T}_3(k,k_2,k+k_2-k_1){\rm I}_3(k,k_1,k_2)\right]\right\}.
\end{eqnarray}
That is to say, the linearized collision terms may now be written
\begin{equation}
{\rm L}_\pm^\prime={\bf k}.{\bf C}\left\{-\epsilon^2\int_0^\infty dk_1
\left[{\rm F}_1^\prime(k,k_1)\,h(k)+{\rm F}_2^\prime(k,k_1)\,h(k_1)\right]\right\},
\label{compactprime}
\end{equation}
where 
\begin{equation}
{\rm F}_1^\prime(k,k_1)\equiv {\rm F}_1(k,k_1), \quad {\rm
F}_2^\prime(k,k_1)\equiv 3\left({\rm F}_2^b(k,k_1)-{\rm
F}_2^a(k,k_1)\right),
\end{equation}
and ${\rm F}_1(k,k_1)$, ${\rm F}_2^a(k,k_1)$ and ${\rm F}_2^b(k,k_1)$
are given by equations (\ref{F1poly}), (\ref{F2apoly}) , and
(\ref{F2bpoly}) respectively.
\section{Longitudinal Crossover Equation}
\label{App:Longcross}
As discussed in section \ref{sect:hydro}, the thermal conductivity is
a function of the dimensionless parameter, $r\propto B^2/\epsilon^4$,
which controls the ratio of the typical cyclotron frequency for a
thermal carrier to the inelastic scattering rate. The longitudinal
distribution function satisfies the integral equation (\ref{genpsieq})
\begin{equation}
\psi_k=\psi_k^\infty-\left(\frac{\epsilon^2}{QB\hbar
c}\right)^2\varepsilon_k {\rm M}_{k,k_1} \varepsilon_{k_1}{\rm
M}_{k_1,k_2}^\prime\psi_{k_2},
\end{equation}
where $\psi_k^\infty$, ${\rm M}$ and ${\rm M}^\prime$ are given by
equations (\ref{psikinf}) and (\ref{Mdefs}) respectively. For both
numerical and analytic purposes it is convenient to recast this
equation in terms of the dimensionless momenta, $\bar k\equiv
ck/k_BT$. This rescaling modifies the coefficient of $\psi_k^\infty$
given in equation (\ref{psikinf}), and it is convenient to introduce
\begin{equation}
\psi_k \equiv
\frac{\epsilon^2}{\hbar}\left(\frac{k_BT}{QBc}\right)^2\Psi(\bar k),
\end{equation}
and similarly for $\psi_k^\infty$. Adopting this rescaling the
integral equation (\ref{genpsieq}) may be recast in the dimensionless
form
\begin{equation}
\Psi(\bar k)=\Psi_\infty(\bar k)-r^{-1}\int_0^\infty d{\bar k}_1[{\rm
Q}_1(\bar k,\bar k_1)\Psi(\bar k)+{\rm Q}_2(\bar k,\bar k_1)\Psi(\bar
k_1)],
\label{dimcross}
\end{equation}
where 
\begin{equation}
r\equiv \frac{1}{\epsilon^4}\left(\frac{QB}{\hbar}\right)^2
\left(\frac{\hbar c}{k_BT}\right)^4,
\end{equation}
and
\begin{equation}
\Psi_\infty(\bar k)\equiv \bar k\int_0^\infty d\bar k_1
\left[\frac{\bar k e^{\bar k}}{(e^{\bar k}-1)^2}\Phi_1(\bar k,\bar
k_1)+\Phi_2(\bar k,\bar k_1)\frac{\bar k_1e^{{\bar k}_1}}{(e^{\bar
k_1}-1)^2}\right]
\label{dlesspsinf}
\end{equation}
coincides with our previous distribution function (\ref{gfunc}). The
non-trivial kernels, ${\rm Q}_1(\bar k,\bar k_1)$ and ${\rm Q}_2(\bar
k,\bar k_1)$, are given by the integrals
\begin{equation}
{\rm Q}_1(\bar k,\bar k_1)\equiv \int_0^\infty d{\bar k}_2\,{\bar
k}\Phi_1(\bar k,\bar k_1){\bar k}\Phi_1^\prime(\bar k,\bar k_2),
\label{Q1def}
\end{equation}
and
\begin{equation}
{\rm Q}_2(\bar k,\bar k_1)\equiv \int_0^\infty d\bar k_2 \left[{\bar
k}\Phi_1(\bar k,\bar k_2){\bar k} \Phi_2^\prime(\bar k,\bar k_1)
+{\bar k}\Phi_2(\bar k,\bar k_1){\bar k_1}\Phi_1^\prime(\bar k_1,\bar
k_2)+ {\bar k}\Phi_2(\bar k,\bar k_2){\bar k_2}\Phi_2^\prime(\bar
k_2,\bar k_1)\right],
\label{Q2def}
\end{equation}
where ${\rm F}_1(k,k_1)\equiv \Phi_1(\bar k,\bar k_1)$, ${\rm
F}_2(k,k_1)\equiv\Phi_2(\bar k,\bar k_1)$, and their primed
counterparts, are the dimensionless kernels given in Appendix
\ref{App:Poly}. Note that in deriving the results (\ref{dlesspsinf}),
(\ref{Q1def}) and (\ref{Q2def}), we have also taken the massless limit
in accordance with the epsilon expansion.

It is evident from the longitudinal crossover equation
(\ref{dimcross}), that the distribution function interpolates between
$\Psi_\infty(\bar k)$, as $r\rightarrow \infty$, and a zero mode of
the integral operator in the hydrodynamic limit, $r\rightarrow 0$. In
order to see this more formally it is natural to consider an expansion
of the form
\begin{equation}
\Psi(\bar k)=\Psi_0(\bar k)+r\Psi_1(\bar k)+r^2\Psi_2(\bar k)+\dots
\label{knudexp}
\end{equation}
Substituting this expansion into equation (\ref{dimcross}) and
equating coefficients at order $r^{-1}$ one finds that for a
non-vanishing $\Psi_0(\bar k)$ to be present it must be an {\em exact}
zero mode of the integral operator:
\begin{equation}
\int_0^\infty d{\bar k}_1[{\rm Q}_1(\bar k,\bar k_1)\Psi_0(\bar
k)+{\rm Q}_2(\bar k,\bar k_1)\Psi_0(\bar k_1)]=0.
\label{intzm}
\end{equation}
As discussed in section \ref{sect:hydro}, the function
\begin{equation}
\Psi_0(\bar k)\propto \frac{\partial f_0}{\partial\varepsilon_k}\propto 
\frac{e^{\bar k}}{(e^{\bar k}-1)^2}\equiv {\mathcal R}_0(\bar k),
\end{equation}
satisfies this homogeneous condition. Although it will not concern us
here, equating coefficients at higher order in $r$ leads to a
recursive hierarchy of integral equations for the functions
$\Psi_l(\bar k)$. These are similar in spirit (if not in details) to
those encountered in the Knudsen expansion of the Boltzmann equation;
see for example \S 6.2 of the book by Harris.\cite{Harris:BE} Equation
(\ref{intzm}) is particularly important from a numerical perspective,
since any regularization which lifts this zero mode property
(e.g. through truncation of the integration limits or through rounding
errors) may potentially yield a solution starting at order $r$,
instead of order $r^0$, as evident from equation (\ref{knudexp}); on
dimensional grounds this would yield a thermal conductivity of the
Wiedemann--Franz form. In order to recover the exact non-vanishing
hydrodynamic limit (\ref{duality}), it is essential that this zero
mode feature is properly implemented. To this end, let us parameterize
our solutions to the integral equation in the form
\begin{equation}
\Psi(\bar k)\equiv {\mathcal R}_0(\bar k)\Phi(\bar k),
\end{equation}
where $\Phi(\bar k)$ is the solution to be determined. Assisted by the
exponential decay of the prefactor, we convert the integral equation
(\ref{dimcross}) into an approximate matrix equation by expanding
$\Phi(\bar k)$ in a basis of $N_b$ basis functions:
\begin{equation}
\Phi(\bar k)=\sum_{n=0}^{N_b-1}c_n f_n(\bar k).
\end{equation}
Substituting this decomposition into equation (\ref{dimcross}), multiplying by 
${\bar k}^{2}f_m(\bar k)$ and integrating (where we incorporate an extra 
factor of ${\bar k}^2$ for convergence purposes) one obtains the 
matrix equation
\begin{equation}
{\rm A}_{mn}c_n={\rm B}_m,
\end{equation}
where, ${\rm A}_{mn}\equiv {\rm A}_{mn}^{\rm I}+r^{-1}{\rm
A}_{mn}^{\rm II}$, with explicit matrix elements
\begin{equation}
{\rm A}_{mn}^{\rm I}\equiv \int_0^\infty d\bar k\,{\bar k}^{2}
{\mathcal R}_0(\bar k)f_m(\bar k)f_n(\bar k),
\end{equation}
and
\begin{equation}
{\rm A}_{mn}^{\rm II}\equiv \int_0^\infty d\bar kd\bar k_1d\bar k_2\,
\bar k^{2}f_m(\bar k)\left[{\mathcal R}_0(\bar k)f_n(\bar k){\rm
Q}_1^{({\rm A})}(\bar k,\bar k_1,\bar k_2)+{\rm Q}_2^{(\rm A)}(\bar
k,\bar k_1,\bar k_2) {\mathcal R}_0(\bar k_1)f_n(\bar k_1)\right],
\end{equation}
where ${\rm Q}_i^{(\rm A)}(\bar k,\bar k_1,\bar k_2)$ denote the
integrands (or arguments) of the integral representations
(\ref{Q1def}) and (\ref{Q2def}). In addition
\begin{equation}
{\rm B}_m\equiv \int_0^\infty d\bar k d\bar k_1\,
{\bar k}^{3}f_m(\bar k)\left[\bar k{\mathcal R}_0(\bar k)
\Phi_1(\bar k,\bar k_1)+
\Phi_2(\bar k,\bar k_1)\bar k_1{\mathcal R}_0(\bar k_1)\right].
\end{equation}
The dimensionless thermal conductivity parameter is defined as
\begin{equation}
g\equiv \frac{1}{3\pi^2}\int_0^\infty d\bar k \bar k^4\Psi(\bar k),
\end{equation}
and is thus approximated as
\begin{equation}
g\simeq \frac{1}{3\pi^2}\sum_{n=0}^{N_b-1}c_n\int_0^\infty d\bar k\,\bar k^{4}
{\mathcal R}_0(\bar k)f_n(\bar k).
\end{equation}
Evaluating the matrix elements numerically (whilst implementing the
zero mode condition ${\rm A}_{m0}^{\rm II}=0$ exactly) we may solve
the linear system of equations for the coefficients $c_n$. Using a
basis of $N_b=7$ monomials $(1,k,\dots,k^6)$ we plot $g$ as a function
of $r$ in Fig.~\ref{Fig:gfactorevol}. It is readily seen that within
the numerical accuracy of our Monte Carlo integrations (approximately
$3\%$) this dimensionless coefficient interpolates between the result
reported in our Letter,\cite{Bhaseen:magneto} and the hydrodynamic
result,\cite{Hartnoll:Nernst} as the parameter $r\propto
B^2/\epsilon^4$ is varied.


\end{document}